\documentclass[aps,prd,twocolumn,amsmath,amssymb,amsfonts,nofootinbib,superscriptaddress]{revtex4-1}
\usepackage[utf8]{inputenc}
\usepackage{graphicx}
\usepackage{subfigure}
\usepackage{datatool}
\usepackage[scientific-notation=true]{siunitx}
\usepackage[dvipsnames]{xcolor}
\usepackage{lineno}
\usepackage{xspace}
\usepackage{orcidlink}
\usepackage{url}

\usepackage{breakurl}
\usepackage{lineno}
\usepackage{longtable}
\usepackage[shortlabels]{enumitem}
\usepackage[capitalise]{cleveref}

\def\gwh{GW\xspace}
\def\cwh{continuous-wave\xspace}
\def\cws{continuous waves\xspace}
\def\gws{GWs\xspace}

\def\dm{DM\xspace}

\def\pbh{PBH\xspace}
\def\pbhs{PBHs\xspace}

\def\fh{frequency-Hough\xspace}
\def\lvk{LIGO, Virgo and KAGRA\xspace}

\def\pn{PN\xspace}
\def\psd{PSD\xspace}

\def\GFH{Generalized frequency-Hough\xspace}

\def\snr{signal-to-noise ratio\xspace}

\newcommand{\bea}{\begin{align}}
\newcommand{\eea}{\end{align}}
\newcommand{\be}{\begin{equation}}
\newcommand{\ee}{\end{equation}}

\newcommand{\TFFT}{T_\text{FFT}}

\newcommand{\Tobs}{T_\text{PM}}

\newcommand{\Mc}{\mathcal{M}}

\newcommand{\fmin}{f_\text{min}}
\newcommand{\fmax}{f_\text{max}}
\newcommand{\fgw}{f_\text{gw}}
\newcommand{\fdot}{\dot{f}_\text{gw}}
\newcommand{\ssm}{sub-solar mass\xspace}
\newcommand{\msun}{\ensuremath{M_\odot}\xspace}

\def\erfc{\mathrm{erfc}}

\graphicspath{{./figures/}}

\begin{document}

\title{Method to search for inspiraling planetary-mass ultra-compact binaries using the generalized frequency-Hough transform in LIGO O3a data}

\author{Andrew L. Miller\,\orcidlink{0000-0002-4890-7627}}
\email{andrew.miller@nikhef.nl}
\affiliation{Nikhef -- National Institute for Subatomic Physics,
Science Park 105, 1098 XG Amsterdam, The Netherlands}
\affiliation{Institute for Gravitational and Subatomic Physics (GRASP),
Utrecht University, Princetonplein 1, 3584 CC Utrecht, The Netherlands}
\author{Nancy Aggarwal}
\email{nqaggarwal@ucdavis.edu}
\affiliation{University of California, Davis, Department of Physics,
Davis, CA 95616, USA}
\author{Sébastien Clesse}
\affiliation{Service de Physique Th\'eorique, Universit\'e Libre de Bruxelles, Boulevard du Triomphe CP225, B-1050 Brussels, Belgium}
\author{Federico De Lillo}
\affiliation{Université catholique de Louvain, B-1348 Louvain-la-Neuve, Belgium}
\author{Surabhi Sachdev}
\affiliation{School of Physics, Georgia Institute of Technology, Atlanta, GA 30332, USA}
\author{Pia Astone}
\affiliation{INFN, Sezione di Roma, I-00185 Roma, Italy}
\author{Cristiano Palomba}
\affiliation{INFN, Sezione di Roma, I-00185 Roma, Italy}
\author{Ornella J. Piccinni}
\affiliation{OzGrav, Australian National University, Canberra, Australian Capital Territory 0200, Australia}
\author{Lorenzo Pierini}
\affiliation{INFN, Sezione di Roma, I-00185 Roma, Italy}

\date{\today}

\begin{abstract}

Gravitational waves from sub-solar mass primordial black holes could be detected in LIGO, Virgo and KAGRA data. Here, we apply a method originally designed to look for rapidly spinning-down neutron stars, the generalized frequency-Hough transform, to search for planetary-mass primordial black holes using data from the first half of the third observing run of advanced LIGO. In this companion paper to \cite{Miller:2024fpo}, in which the main results of our search are presented, we delve into the details of the search methodology, the choices we have made regarding the parameter space to explore, the follow-up procedure we use to confirm or reject possible candidates returned in our search, and a comparison of our analytic procedure of generating upper limits to those obtained through injections. 


\end{abstract}

\maketitle


\section{Introduction}

The nature of dark matter (DM) is one of the most perplexing problems in contemporary physics. Despite decades of research, \dm has managed to elude detection. Therefore, a multi-pronged worldwide research program is afoot to find dark matter, focusing on various candidates, such as weakly interacting massive particles, sterile neutrinos, axions, scalar fields, and primordial black holes (PBHs)    \cite{BERTONE2005279,bertone2018new,KLASEN20151}. 

PBHs, a  dark matter candidate, could have formed in the early universe from the collapse of density fluctuations \cite{Carr:2016drx}. Since PBHs do not have stellar origins, their mass function is unrelated to the stellar mass function, allowing for the possibility that PBHs are lighter than a solar mass. If BHs lighter than a solar mass are discovered, this would be an almost unambiguous proof of their primordial origin, in contrast to the BHs above a solar mass observed so far, whose origins could be astrophysical or primordial \cite{LISACosmologyWorkingGroup:2023njw,Yamamoto:2023tsr}

Constraints across the possible \pbh mass range can be derived through the lack of imprints in the Cosmic Microwave Background (CMB) of radiation emitted through gas accretion onto \pbhs \cite{carr:1981xxx,Ricotti:2007au} or the lack of x-ray observations of accretion of interstellar gas onto \pbhs \cite{Gaggero:2016dpq}, the lack of x-rays and gamma rays predicted from Hawking evaporation \cite{Page:1976wx}, e.g. \cite{Boudaud:2018hqb}, and the lack of disruption of stellar clusters due to dynamical effects on velocities of stars within those systems \cite{Brandt:2016aco,Koushiappas:2017chw}. However, actually \emph{searching} for \pbhs is another story: traditional searches for PBHs use gravitational lensing of distant galaxies or supernovas, although, no searches so far have not discovered a PBH candidate \cite{tisserand2007limits,niikura2019constraints,Croon:2020ouk}, and it turns out that searches for \gws from \pbhs may be able to provide competitive or even stronger constraints at certain \pbh masses \cite{Arvanitaki:2014faa,Green:2020jor}.

The detection of GWs from black holes much heavier than a few solar masses \cite{Abbott:2016blz} by \lvk, previously assumed to be non-existent, has motivated a deeper scrutiny of the PBH DM scenario \cite{Sasaki:2016jop,Clesse:2016vqa,Bird:2016dcv} and an interest in constraining PBHs using a diverse set of observations. GWs from PBHs in binaries provide one such independent way to search for or constrain PBHs. 

Often, a matched filtering algorithm is used to search for GWs from sources whose phase evolution we exactly know, e.g. from the inspiral, merger, and ringdown of binary black holes (BBH) \cite{Allen:2005fk}, or from known isolated pulsars \cite{Dupuis:2005xv}. In the case of binary mergers, this requires generating signal templates using the post-Newtonian (PN) approximation for the inspiral, numerical relativity simulations for the merger, and BH perturbation theory for the ringdown, across a 15-dimensional space of various BBH parameters (masses, spins, etc.) \cite{maggiore2008gravitational}. The template duration increases with decreasing black hole mass, making such a method computationally unfeasible for very light BBHs \cite{Miller:2024rca}. Therefore, matched filtering algorithms are used to search for BBHs of mass \( \sim 0.1\, \mathrm{M}_\odot\) and above \cite{LIGOScientific:2021job,Phukon:2021cus,Nitz:2021mzz,Nitz:2022ltl,Bandopadhyay:2022tbi,LIGOScientific:2022hai}.

While the template duration at low masses can be really long, the frequency evolution is expected to be somewhat slow in the audio band, compared to higher-mass systems, and the systems will merge well outside the frequency bands of the \lvk detectors. Thus, we can apply methods that are designed to search for \cws, quasi-monochromatic \gws canonically from isolated neutron stars, to look for the slow inspiral of a binary \pbh system. Such \cwh methods were created because the matched filter was too computationally expense to explore the full parameter space of isolated neutron stars -- sky position, frequency, rate of change of frequency, etc.--; thus, \cwh methods would also allow us to avoid using template banks that comprise the bulk of computational cost of matched filtering, with a small trade-off in sensitivity that could be mitigated in a real search \cite{Miller:2023rnn}.


We previously devised a way to search for asteroid- and planetary-mass \pbh binaries \cite{Miller:2021knj} that was based on the \fh transform \cite{Astone:2014esa}. Its primary limitation was that it could not be applied to \pbh inspirals in which the \gwh frequency was changing faster than $10^{-9}$ Hz/s. In this work, we present a method, based on the \GFH Transform \cite{Miller:2018rbg,Miller:2020kmv}, with which we can search for \gwh signals that would last longer than durations that matched filtering can handle, and whose frequencies would change more rapidly than frequency drifts that standard CW searches can handle. We apply this method to data from the first half of the third observing run of Advanced LIGO, search for candidates, and in the absence of any viable \pbh candidates, set upper limits on them \cite{Miller:2024fpo}. This method is applicable for \pbh binaries across a wide range of masses, in both equal-mass and highly asymmetric mass-ratio binaries, and is one of a few that has been recently developed to tackle this problem \cite{Andrés-Carcasona:2023df,Alestas:2024ubs}.

This paper goes into the details of the search methodology and accompanies another paper that presents the search results \cite{Miller:2024fpo}. This paper is organized as follows: in Section II, we describe the \gwh inspiral signal from two masses in orbit around each other. In Section III, we explain our algorithm, specifically transforming the data into frequency domain and later into the Hough plane using the \GFH transform. We also show the parameter space over which this method is applicable and describe the configurations over which the search is conducted to fully cover the parameter space in minimal computation time. In Section IV, we detail the procedure to follow up and veto above-threshold candidates returned by the search that were due to noise. We find that no candidates survive the veto procedure. In Section V, we devise a method to calculate upper limits from the search results on the existence of PBHs in the nearby universe. In Section VI, we calculate the computational cost of our search, and compare it to \cwh and matched-filtering searches for \pbhs before concluding in section VII.

\section{Signal model}

Two compact objects in orbit around their center of mass will emit \gws as they approach each other. Equating the orbital energy loss with \gwh power, we can obtain the rate of change of the frequency over time, i.e. the spin-up $\fdot$, in the quasi-Newtonian limit (i.e. far from merger) \cite{maggiore2008gravitational}:

\begin{align}
    \dot{f}_{\rm gw}&=\frac{96}{5}\pi^{8/3}\left(\frac{G\Mc}{c^3}\right)^{5/3} f_{\rm gw}^{11/3}\equiv k f_{\rm gw}^{11/3} \nonumber \\
    &\simeq 1.25\times 10^{-4}\text{ Hz/s} \left(\frac{\Mc}{10^{-3}M_\odot}\right)^{5/3}\left(\frac{f_{\rm gw}}{100\text{ Hz}}\right)^{11/3},
    \label{eqn:fdot_chirp}
\end{align}
where $\Mc\equiv\frac{(m_1m_2)^{3/5}}{(m_1+m_2)^{1/5}}$ is the chirp mass of the system, $f_{\rm gw}$ is the \gwh frequency, $c$ is the speed of light, and $G$ is Newton's gravitational constant.

To obtain the signal frequency evolution $f_{\rm gw}(t)$ over time, we integrate \cref{eqn:fdot_chirp} with respect to time $t$:
\begin{equation}
f_{\rm gw}(t)=f_0\left[1-\frac{8}{3}kf_0^{8/3}(t-t_0)\right]^{-\frac{3}{8}}~,
\label{eqn:powlaws}
\end{equation}
where $t_0$ is a reference time for the \gwh frequency $f_0$. 

The amplitude $h_0(t)$ of the \gwh signal also evolves with time \cite{maggiore2008gravitational}:

\begin{align}
h_0(t)&=\frac{4}{d}\left(\frac{G \Mc}{c^2}\right)^{5/3}\left(\frac{\pi f_{\rm gw}(t)}{c}\right)^{2/3} \nonumber \\
&\simeq 2.56\times 10^{-23}\left(\frac{1\text{ kpc}}{d}\right)\left(\frac{\Mc}{10^{-3}\msun}\right)^{5/3}\left(\frac{f_{\rm gw}}{20\text{ Hz}}\right)^{2/3},
\label{eqn:h0}
\end{align}
where $d$ is the luminosity distance to the source.

Inverting \cref{eqn:powlaws}, we can also write down an expression for the time the signal spends between two frequencies:

\begin{equation}
    \Delta t=-\frac{3}{8}\frac{f_{\rm gw}^{-8/3}-f_0^{-8/3 }}{k},
\end{equation}
which, in the limit that $f_{\rm gw}\rightarrow\infty$, determines the time to merger $t_{\rm merg}$

\begin{align}
    t_{\rm merg}&\simeq\frac{5}{256}\left(\frac{1}{\pi f_0}\right)^{8/3}\left(\frac{c^3}{G\Mc}\right)^{5/3}\nonumber \\
    &\simeq 3.4\text{ days} \left(\frac{100\text{ Hz}}{f_0}\right)^{8/3} \left(\frac{10^{-3}\msun}{\Mc}\right)^{5/3}.
    \label{eqn:tmerg}
\end{align}
Signals whose frequencies follow \cref{eqn:powlaws} and last $\sim$ hours to days in the detector are called ``transient continuous waves'', and would typically have chirp masses of $\mathcal{O}(10^{-5}-10^{-2})\msun$. Other signals that last for durations much longer than a few days, and whose frequency evolution is almost monochromatic (that is, $\dot{f}\lesssim 10^{-9}$ Hz/s), are called ``continuous waves'', and would have chirp masses less than $10^{-5}\msun$.

As seen above, systems with $\Mc\ll\msun$ spend a long time in the detector frequency band before merging relative to those with $\Mc\gtrsim\msun$. 
Such long-duration signals are problematic for conventional matched-filtering algorithms, thus motivating the development of alternative methods to search for such light-mass \pbhs \cite{Miller:2020kmv,Andrés-Carcasona:2023df,Alestas:2024ubs}.

\begin{figure*}
    \centering
    \includegraphics[width=\textwidth]{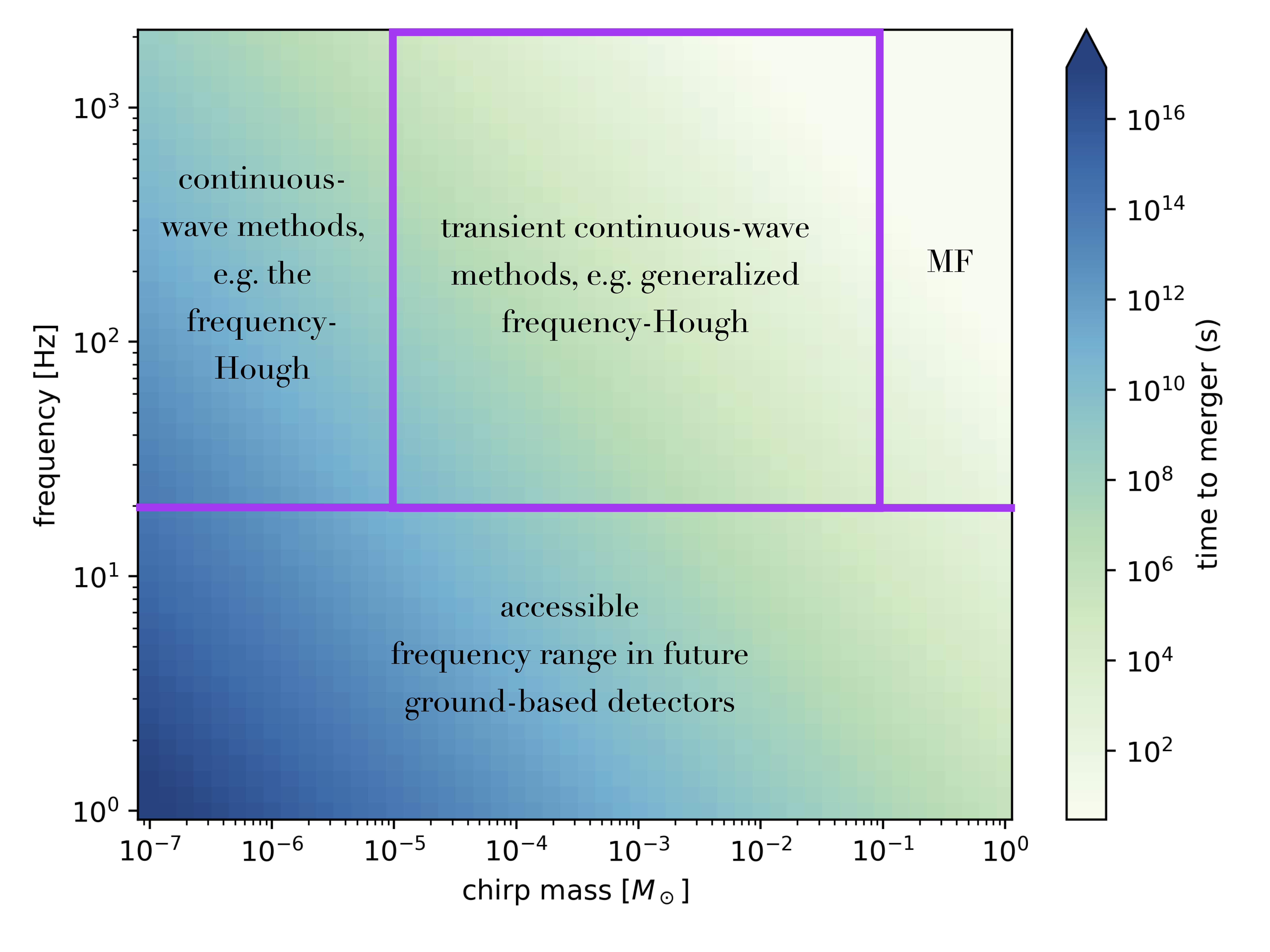}
    \caption{Time to merger as a function of the chirp mass and inspiral \gwh frequency. The plot is divided into different regions depending on which methods could be used to probe \pbhs inspiraling with different chirp masses: MF (matched filtering), transient \cwh methods and traditional \cwh methods. }
    \label{fig:mergtime}
\end{figure*}

In \cref{fig:mergtime}, we plot time to merger as a function of chirp mass and \gwh frequency (\cref{eqn:tmerg}). We divide the figure into different regions depending on which methods could be used to probe particular chirp masses: matched filtering (MF) for $[0.1,1]\msun$ \cite{LIGOScientific:2022hai}, transient \cwh methods, such as the \GFH \cite{Miller:2018rbg,Miller:2020kmv} or pattern-recognition techniques \cite{Sun:2018hmm,Oliver:2018dpt,Miller:2019jtp,Andrés-Carcasona:2023df,Alestas:2024ubs}, for $[10^{-5},10^{-1}]\msun$, and traditional \cwh methods, such as the original \fh \cite{Astone:2014esa}, to probe $[10^{-7},10^{-5}]\msun$. Frequencies between $[1,20]$ Hz could be accessible now in \gwh searches, however the trade-off in the increased computational cost for matched-filtering methods, and marginal sensitivity gained by analyzing longer signals, which will be weaker than at high frequencies (\cref{eqn:h0}), is not worth it \cite{LIGOScientific:2019kan}. Furthermore, due to the relatively poor sensitivity of LIGO at such low frequencies, even the \cwh and transient \cwh methods are not likely to benefit much from analyzing such low-frequency data. However, in future ground-based detectors, such low-frequency information could be critical for sky localization of compact binaries, see e.g. \cite{Cannon:2011vi,Zhao:2017cbb,Chan:2018csa,Sachdev:2020lfd,Banerjee:2022gkv,Nitz:2021pbr,Baral:2023xst,Miller:2023rnn}.

\section{Search implementation details}
\subsection{Creation of time-frequency peakmaps}
Our search of \gwh data for planetary-mass \pbhs begins with data structures called the short fast Fourier transform databases (SFDBs) \cite{Astone:2005fj}, which are sets of 50\% interlaced fast Fourier transforms (FFTs) for the entire observing run of length 1024 s. We start with these data structures, instead of the raw strain data $h(t)$, because a series of cleaning procedures have already been applied to remove random bursts of noise called glitches from the raw strain data $h(t)$ \cite{Acernese:2009zz}. From these FFTs, we actually inverse FFT them to the time-domain, recovering the (cleaned) $h(t)$, and then proceed with our analysis. Over the time $\Tobs$ we would like to analyze data, we take FFTs of length $\TFFT$, which is determined by ensuring that the signal power is confined to one frequency bin in each FFT. In each FFT, we estimate the noise power spectral density (PSD) using an auto-regressive method \cite{Astone:2014esa} that tracks slow variations in the noise while predominately ignoring monochromatic lines. This is important: since $\TFFT$ is chosen such that the signal is confined to one frequency bin, we should reduce the impact of the monochromatic signal on the estimation of the noise \psd in each FFT. 

{After we compute the noise \psd, we calculate the equalized power spectrum $ R_{k,l}$, defined as:}

\begin{equation}
    R_{k,l} = \frac{|\text{FFT}(h(t))|^2}{\text{PSD}}
\end{equation}
{at each time start-time of the FFT $k$ and each frequency $l$.} 

The statistical distribution of $R_{k,l}$ is exponential, with mean and standard deviation both equal to unity. We then select local maxima of $R_{k,l}$ above the threshold $\theta_{\rm thr}=2.5$, chosen based on extensive studies of how injected \gwh signals behave in non-stationary, non-Gaussian noise \cite{Astone:2014esa}. This threshold means that we keep particular time-frequency points whose equalized powers are 2.5 standard deviations away from the mean. Furthermore, we impose the ``local maxima'' criteria because the power due to noise disturbances may spread into adjacent frequency bins of their central frequencies, thus we limit the impact of this effect, spectral leakage, when selecting important time-frequency points.

In \cref{fig:pmhm}, left-hand column, we illustrate taking FFTs of length $\TFFT$ at times labeled $1,2,$ and $3$, and show what the spectra look like in the time-frequency plane. It is in this 3-dimensional plot that we select time-frequency points whose equalized power is above a threshold and also a local maximum.

The output of this part of the pipeline is a time-frequency ``peakmap'', which is essentially a collection of ones, i.e. ``peaks'', at different times and frequencies within $\Tobs$. In \cref{fig:pmhm}, the peakmap is shown in the top right-hand corner with an injected \pbh signal. We note that the presence of white space in this figure represents the time-frequency points whose equalized powers were \emph{not} above $\theta_{\rm thr}$. While the color bar in this plot shows the value of the equalized power, the equalized power does not matter for subsequent stages in the analysis, since all ``peaks'' in the time-frequency map are given values of 1. We disregard the power at each peak to be more robust against non-Gaussian noise disturbances \cite{Astone:2005fj,Astone:2014esa}, since powerful noise lines are given weights of ``1'' at each time they appear, thus reducing background contamination.

\subsection{The Generalized frequency-Hough transform}

The \GFH is a pattern-recognition technique that was originally designed to find rapidly spinning-down neutron stars following an arbitrary power-law \cite{Miller:2018rbg}. It is in fact based on a method, the \fh transform, that was designed to find very slowly spinning down neutron stars anywhere in the sky \cite{Astone:2014esa}. Briefly, the \fh maps points in the time-frequency peakmap to lines in the frequency-spindown plane of the source. 

 The \fh works by summing the peaks in the peakmap along different ``tracks'', in which each track can be described by a line, with $y$-intercept $f_0$ and slope $\dot{f}$. In other words, each track has the form:

\begin{equation}
f_{\rm gw} = f_0+\dot{f}_{\rm gw}(t-t_0).
\label{eqn:linf}
\end{equation}
The input to the \fh is the peakmap, i.e. $t$ and $\fgw$, and $t_0=T_{\rm obs}/2$ is a chosen reference time. Subsequently, for each choice of $f_0$ and $\dot{f}_{\rm gw}$, a line can be drawn in the time-frequency peakmap, and the peaks along that line are summed. The number of peaks that fall along that track are histogrammed in the two-dimensional $f_0$-$\dot{f}$ plane, also called the Hough map. The \fh is sensitive to inspiraling \pbh systems with $\Mc\lesssim 10^{-5}\msun$, i.e. to slowly spinning up binaries. In fact, to arrive at \cref{eqn:linf}, one can binomially expand \cref{eqn:powlaws}, which corresponds to $\dot{f}_{\rm gw}(t-t_0)\ll f_0$.

\begin{figure*}
    \centering
    \includegraphics[width=0.99\textwidth]{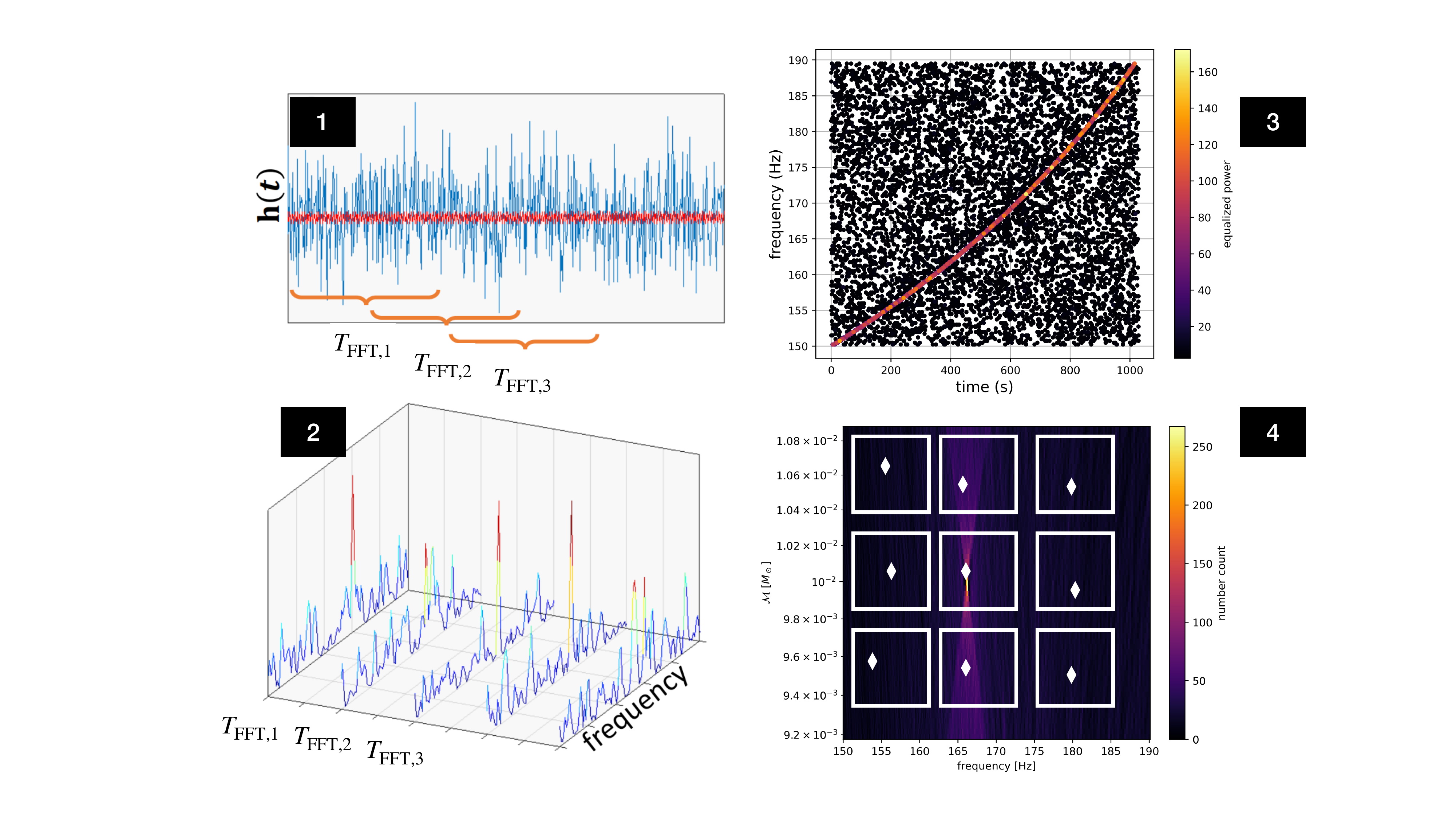}
    \caption{Top left: raw strain data $h(t)$, in which we show how we divide the data set into overlapping chunks of length $\TFFT$. We highlight the (weak) signal that is simulated in this time-domain data in red. Bottom left: a 3-D plot showing the result of taking a series of 50\% interlaced FFTs over $\Tobs$. We see that a clear peak in red at each time, since the signal is monochromatic in each FFT. Top right: the peakmap containing a simulated \gwh signal from a \pbh inspiral with a chirp mass of $10^{-2}M_\odot$ in white noise. Each ``column'', that is, each time, corresponds to one FFT taken of a stretch of data of a given length $\TFFT$. Within each $\TFFT$, the signal is confined to one frequency bin, evidence by the presence of one bright pixel at each time. Bottom right: the result of the \GFH applied to the peakmap. Here, the frequency on the $x$-axis corresponds to the signal frequency at $t_0=\Tobs/2$ in the left-hand figure. Images 1 and 2 are adapted from \cite{lorenzoposter}.}
    \label{fig:pmhm}
\end{figure*}

Now, when the \fh was generalized to follow power laws that describe neutron stars of the form

\begin{equation}
\dot{f}_{\rm gw} = C f_{\rm gw}^n,
\end{equation}
where $C$ is a negative proportionality constant and $n$ is the power of the power-law, whose integral is:

\begin{equation}
f_{\rm gw}(t)=f_0\left[1-(n-1)Cf_0^{n-1}(t-t_0)\right]^{-\frac{1}{n-1}}~,
\label{eqn:genpowlaw}
\end{equation}
it was necessary to rewrite \cref{eqn:genpowlaw} in a form that is linear to apply the \fh, so the following change of coordinates was used:

\begin{equation}
z=1/f_{\rm gw}^{n-1} ; z_0=1/f_0^{n-1}.
\end{equation}
At this point, the plane of the peakmap is no longer $t-f$ but $t-z$, and the lines over which we sum the peaks have the form

\begin{equation}
z=z_0+(n-1)C(t-t_0)
\end{equation}
Specializing to the case of inspiraling \pbhs, with spin up (so $C\rightarrow-k$) and $n=11/3$, we arrive at:

\begin{equation}
z=z_0-\frac{8}{3}k(t-t_0)
\end{equation}
The above is just algebra: while the method has been applied to look for \gws for different physical phenomena, conceptually nothing has changed in terms of how the \fh is implemented. The only additional requirement is that the time-frequency peakmap is first transformed to the $t-z$ plane before applying the \fh, and this transformation to the $t-z$ plane depends on $n$. An example Hough map is shown in \cref{fig:pmhm}, bottom right-hand corner, which contains the result of applying the \GFH to the peakmap in the top right-hand corner. Different tracks in the peakmap correspond to different $f_0-\Mc$ pairs, and the number of peaks that fall along each track is histogrammed in the Hough map in \cref{fig:pmhm}.



\subsection{Search configurations}\label{app:config}


Because we have limited computational power, we should determine the range of frequencies and chirp masses to search over in order to maximize our sensitivity to \pbh inspirals. To do this, we consider the equation for the sensitivity of the \GFH \cite{Miller:2020kmv} to inspiraling systems:

\begin{align}
d_{\rm max}&=\left(\frac{G \Mc}{c^2}\right)^{5/3}\left(\frac{\pi}{c}\right)^{2/3} \frac{\TFFT}{\sqrt{\Tobs}}\left(\sum_x^N \frac{f^{4/3}_{\text{gw},x}}{S_n(f_{\text{gw},x})}\right)^{1/2} \nonumber \\ &\times \left(\frac{p_0(1-p_0)}{Np^2_1}\right)^{-1/4}\sqrt{\frac{\theta_{\rm thr}}{\left(CR-\sqrt{2}\erfc^{-1}(2\Gamma)\right)}}.
\label{eqn:dmax}
\end{align} 
Here, $p_0=e^{-\theta_{\rm thr}}-$  $e^{-2\theta_{\rm thr}}$ $+\frac{1}{3}e^{-3\theta_{\rm thr}}$ is the probability of selecting a peak above $\theta_{\rm thr}=2.5$ in noise, $p_1$ = $e^{-\theta_{\rm thr}}-$  2$e^{-2\theta_{\rm thr}}$ $+e^{-3\theta_{\rm thr}}$, $N=\Tobs/\TFFT$ is the number of FFTs used to make the peakmap, and $S_n$ is the detector noise PSD. $x$ indicates the sum over the theoretical frequency track for the chirp mass $\Mc$ with starting frequency $f_{\mathrm{gw},x=0}$, where all frequencies $f_{\mathrm{gw},x}$ are weighted by the noise PSD.

We see here a complicated dependence amongst how long we make the peakmaps $\Tobs$, which FFT length we use $\TFFT$, the PSD $S_n(f_{\text{gw},x})$, which chirp masses we consider $\Mc$, and the frequency evolution of the system over time (given by the sum over $N$ FFTs). Based on this equation, it is in fact not the case that observing for as long as possible will guarantee the best sensitivity, primarily due to the interplay between increasing signal frequency, the changing noise PSD, and the fact that $\dot{f}$ increases with frequency, which then limits the $\TFFT$ we can take.

In practice, we divide the searched parameter space into a set of ``configurations'', where each configuration contains a value for $\fmin,\fmax,\Mc_{\rm min},\Mc_{\rm max},\TFFT$ and $\Tobs$.
For each configuration's frequency and chirp mass range, there is an optimum $\Tobs$ and $\TFFT$, determined through multiple considerations:
\begin{enumerate}[a)]
    \item The signal frequency modulation given by \cref{eqn:powlaws} must be confined to half a frequency bin in each FFT:
\begin{equation}
    \dot{f}_{\rm gw}\TFFT \leq \frac{1}{2\TFFT} \rightarrow \TFFT \leq \frac{1}{\sqrt{2\dot{f}_{\rm gw}}}.\label{eqn:tfft}
\end{equation}
\item  To ensure that we are indeed sensitive to signals up to 3.5\pn, we also require that the difference between our O\pn approximation for the \gwh frequency $f_{\rm gw}$, and the \gwh frequency up to 3.5\pn $(f_{3.5PN})$, is smaller than the frequency bin of the Fourier transform:

\begin{equation}
    |f_{\rm gw}(t)-f_{3.5PN}(t)| \leq \frac{1}{\TFFT},
\end{equation}
where $f_{3.5PN}(t)$ is given by Eq. 5.258 in \cite{maggiore2008gravitational}.

\item The spin-up increases with time as seen in \cref{eqn:fdot_chirp}, which decreases the corresponding allowed $\TFFT$ (\cref{eqn:tfft}). Since the sensitivity of a search increases with signal amplitude, observation time and $\TFFT$ \cite{Miller:2020kmv}, and inversely with the noise \psd,
there should exist an optimal peakmap duration $\Tobs$ and $\TFFT$ that maximize the sensitivity to a particular chirp mass. 
\item We require that the minimum distance reach of our search to be $\sim 0.1$ kpc.
\item To reduce the computational cost of the search \cite{Miller:2020kmv}, we ensure that within $\TFFT$, the signal modulation due to the Doppler motion of the earth with respect to any sky location is confined to one frequency bin for the whole observing run $T_{\rm obs}$. In other words, we do not have consider a grid in the sky to do our search, and are sensitive to sources coming from any point in the sky, though we cannot localize where they come from.
\end{enumerate}



With these five criteria in mind, for each chirp mass between $\sim [10^{-5},10^{-2}]\msun$, we vary the possible frequency range $[\fmin,\fmax]$ over which we would look for a signal, which fixes $\Tobs$ and $\TFFT$ by \cref{eqn:tmerg} and \cref{eqn:tfft}, respectively. With these quantities and an estimate for the noise PSD of the detector, we can compute $d_{\rm max}$ in \cref{eqn:dmax}. After looping over values for $\fmin,\fmax,\Tobs,$ and $\TFFT$, for a fixed $\Mc$, we arrive at a set of these parameters that maximizes $d_{\rm max}$. Fig. 8 in \cite{Miller:2020kmv} shows the optimal $\TFFT$ and $\Tobs$ for particular $f_0-\Mc$ pairs.

It would be ideal to create one peakmap per \pbh system with a particular $f_0-\Mc$ and run the \GFH on it. However, that would require an immense amount of computational power. Instead, we know that each peakmap may contain a variety of \pbh signals, even if not optimally sensitive to all of them. Thus, we overlap peakmaps in frequency and time to ensure that, for any given \pbh signal, we would not lose more than 10\% in sensitivity if the signal falls between peakmaps. By applying these selection criteria, we obtain $N_{\rm config}=129$ distinct configurations to search over, the details of which are given as a .txt file \cite{miller_2024_10724845}.



\begin{figure}
    \centering
    \includegraphics[width=0.49\textwidth]{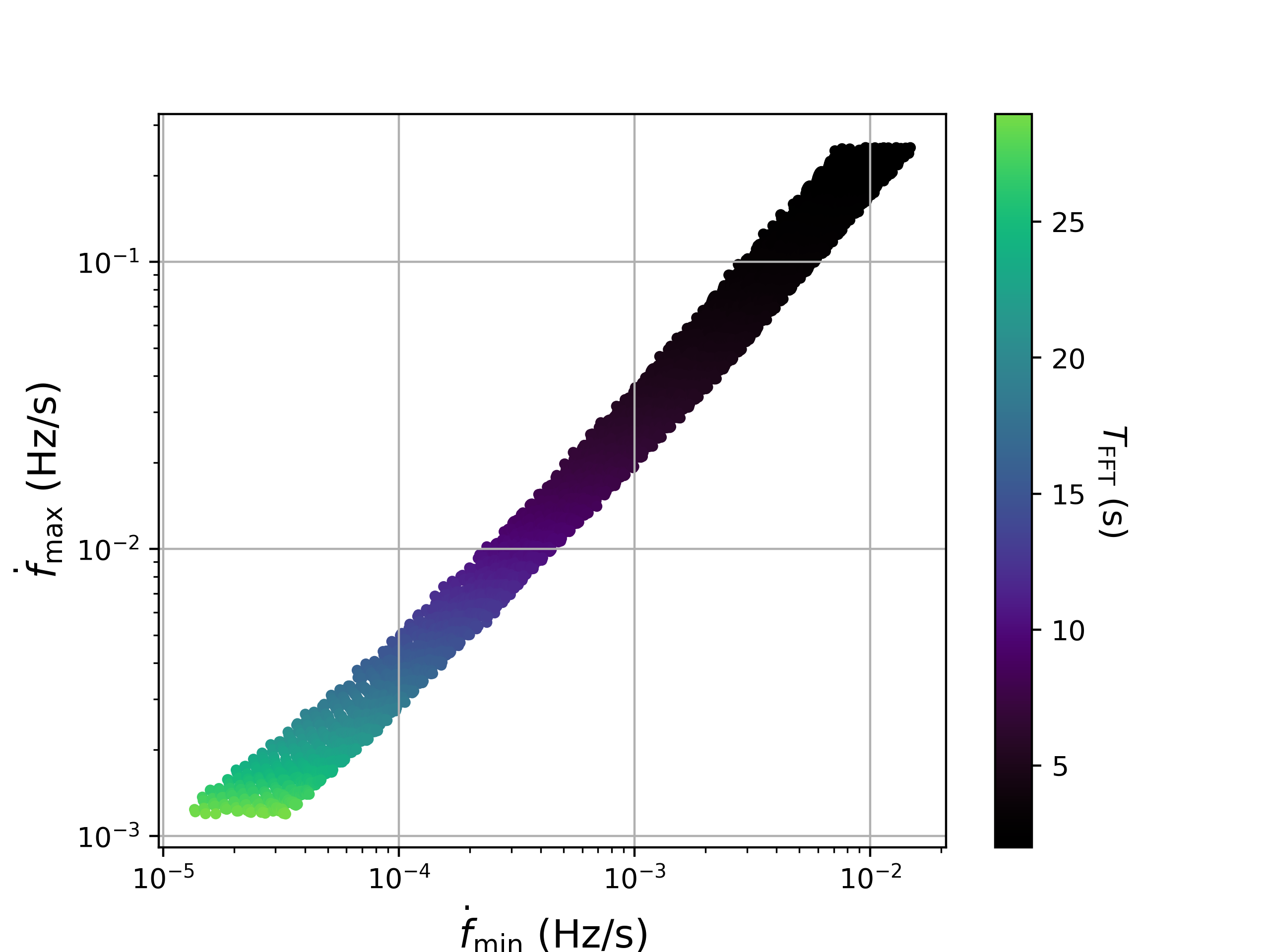}
    \caption{Range of spin-ups considered in \cite{Miller:2024fpo}, with $\TFFT$ colored.}
    \label{fig:fdotmax_vs_fdotmin}
\end{figure}
\begin{figure}
    \centering
    \includegraphics[width=0.49\textwidth]{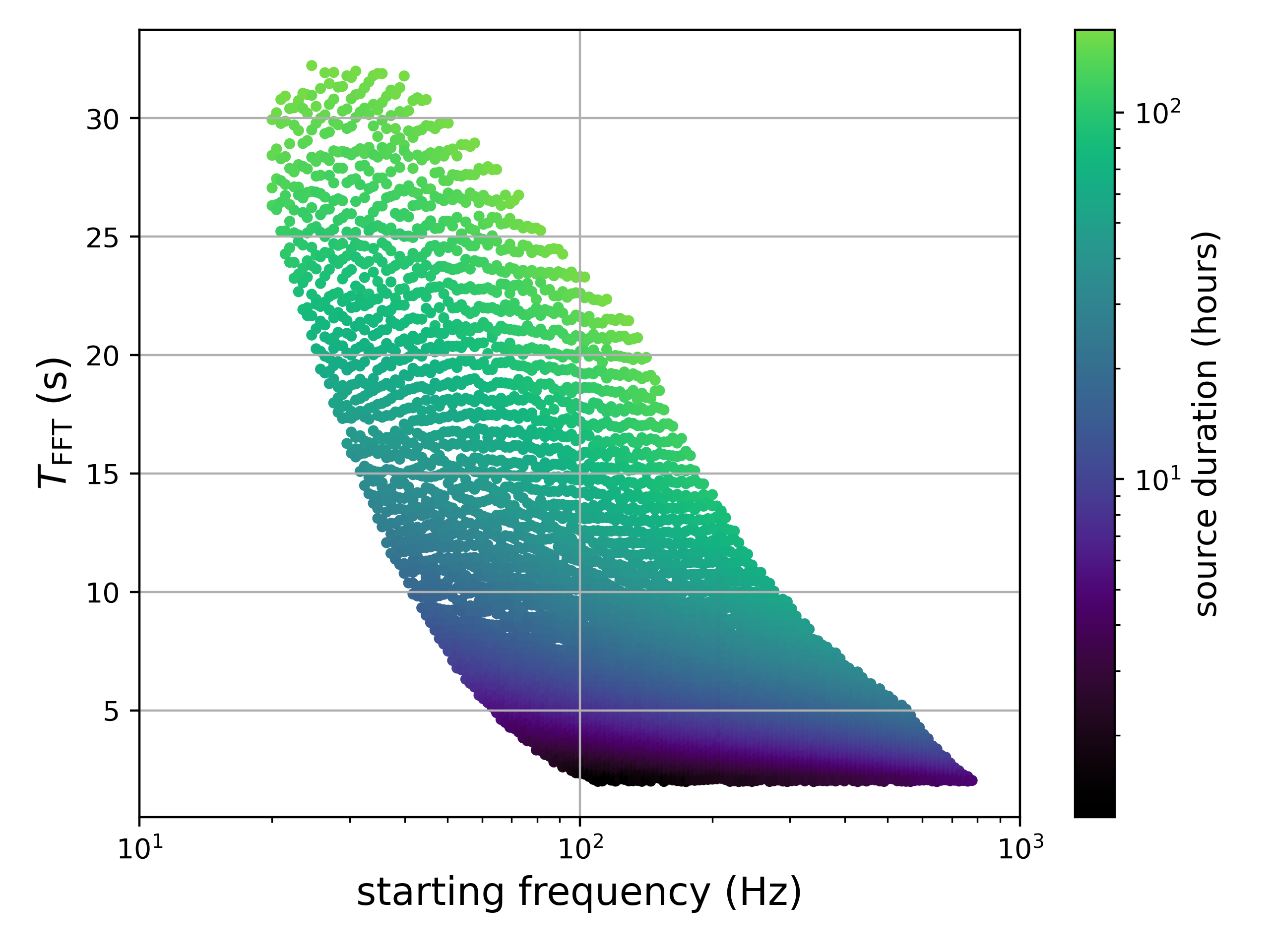}
    \caption{How $\TFFT$ changes as a function of the frequency $f_0$ and $\Tobs$, for chirp mass systems roughly between $[10^{-5},10^{-2}]M_\odot$ }
    \label{fig:tfft-vs-freq-durcol}
\end{figure}

In \cref{fig:fdotmax_vs_fdotmin}, we give the reader an idea of how $\TFFT$ in general changes as a function of the range of spin-ups to which we are sensitive. As indicated in \cref{eqn:tfft}, the slower the inspiral frequency changes, the longer the $\TFFT$ can be. Furthermore, we show in \cref{fig:tfft-vs-freq-durcol} the relationship between $\Tobs$, $\TFFT$ and the starting \gwh frequency of the source. We can see that if we observe the signal at higher frequencies, we have less time to accumulate signal-to-noise ratio, and that its frequency is changing much more rapidly than at lower frequencies, evidenced by the relatively smaller $\TFFT$ length.

\section{Follow-up of candidates returned in the search}

We run the search in each detector's data separately, and select the top 0.1\% candidates in each of the Hough maps in various ``squares'' of frequency/chirp mass, as shown in \cref{fig:pmhm}, bottom right-hand plot. We essentially select the top candidates per white square because in \cite{Miller:2018rbg}, we have found that, even in Gaussian noise, the distribution of peaks and number counts in the peakmap and Hough map, respectively, are skewed towards lower values of $z$. This occurs because the change of coordinates takes peaks that are uniformly distributed in frequency and concentrates them at higher frequencies (lower $z$ values), and
spreads them out at lower frequencies (high $z$ values). Therefore, when the \GFH is performed on the transformed peakmap, more counts tend to accumulate at low $z$ (high frequency) values. By selecting candidates in different squares, whose sizes are determined by the total number of candidates we would like to select in each Hough map, we are able to cover the $z-k$ parameter space in a more uniform manner than if simply selecting the maxima over the whole map. Selecting candidates square by square also has the benefit that we avoid being blinded by disturbances that may appear in the Hough map.

We then perform coincidences between the candidates returned from each detector's data. A ``coincidence'' occurs if the Pythagorean distance between the returned $z_0$ and $k$ ($f_0$ and $\Mc$) of candidates from each detector is less than 3 bins apart \cite{Miller:2018rbg} at the same reference time $t_0$, that is:

\begin{equation}
    \text{dist} = \sqrt{\left(\frac{k_{\rm LHO} - k_{\rm LLO}}{\delta k}\right)^2 + \left(\frac{z_{\rm LHO} - z_{\rm LLO} }{\delta z}\right)^2}
    \label{eqn:dbin}
\end{equation}
where $\delta k$ and $\delta z$ are the bin sizes in each of the coordinates. $\delta k$ varies as a function of $\TFFT$, $k$ ($\Mc$) and $f_0$, while $\delta z$ depends solely on $f_0$ and $\TFFT$ \cite{Miller:2018rbg}. 

After running the search, we have $\sim$10 million coincident candidates within 3 bins of each other, denoted $N_{\rm cand}$, irrespective of their significance, which is addressed in the following steps.

For each of these candidates, we compute a value for our detection statistic, called the critical ratio (CR), in each ``square'' of the Hough map:

\begin{equation}
    CR = \frac{m - \mu }{\sigma}
\end{equation}
where $m$ is the number count in a given $z_0-k$ bin, and $\mu$ and $\sigma$ are the average and standard deviation of the number counts of the bins in the ``square'' around the selected candidate, respectively. The CR is essentially the number of standard deviations a particular candidate is away from the mean of each ``square''.

\begin{figure*}[ht!]
     \begin{center}
        \subfigure[ ]{%
            \label{fig:nkpt}

        \includegraphics[width=0.5\textwidth]{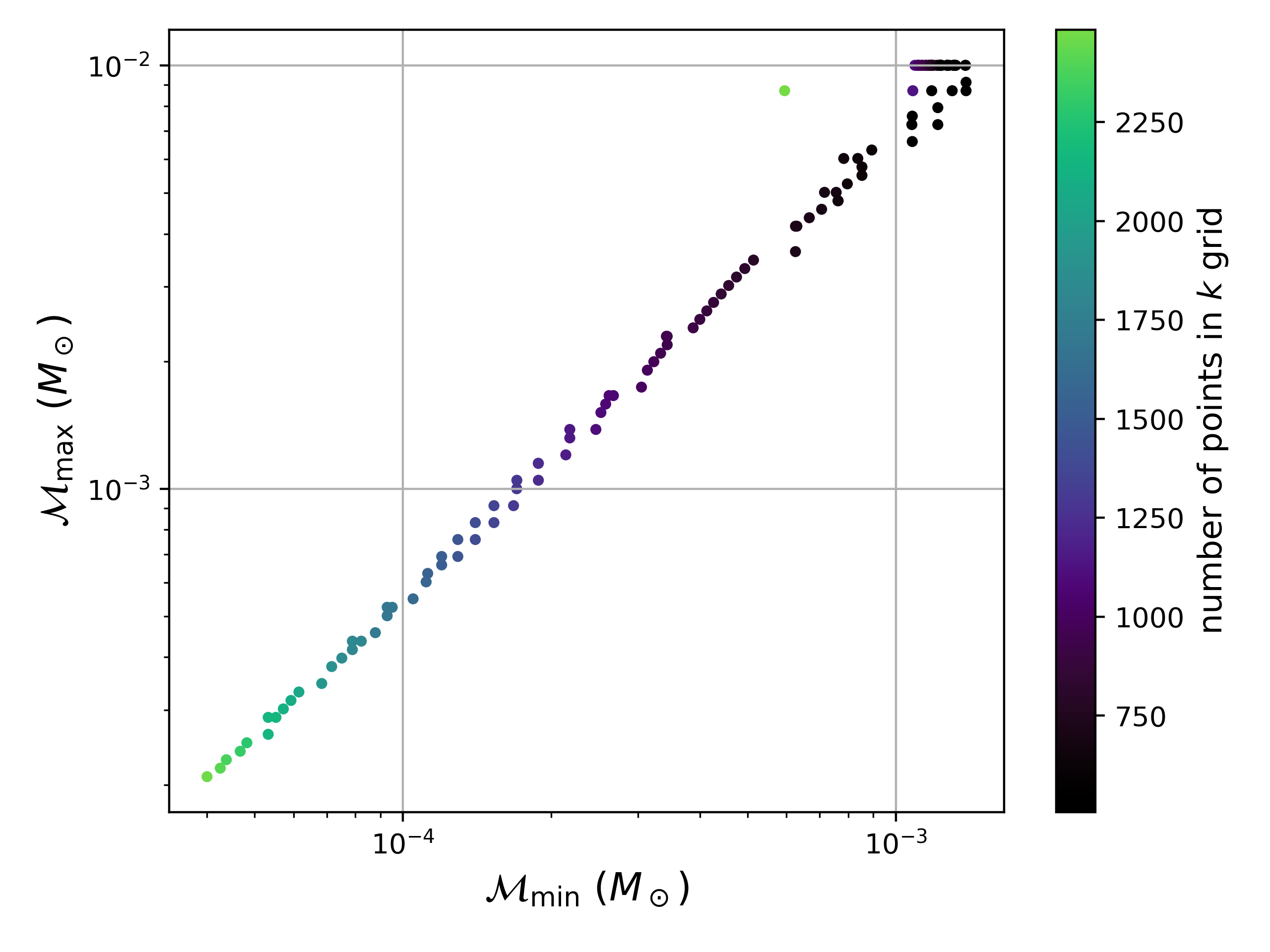}
        }%
        \subfigure[]{%
           \label{fig:n_z_pts_fmin_fmax_parmspace}
           \includegraphics[width=0.5\textwidth]{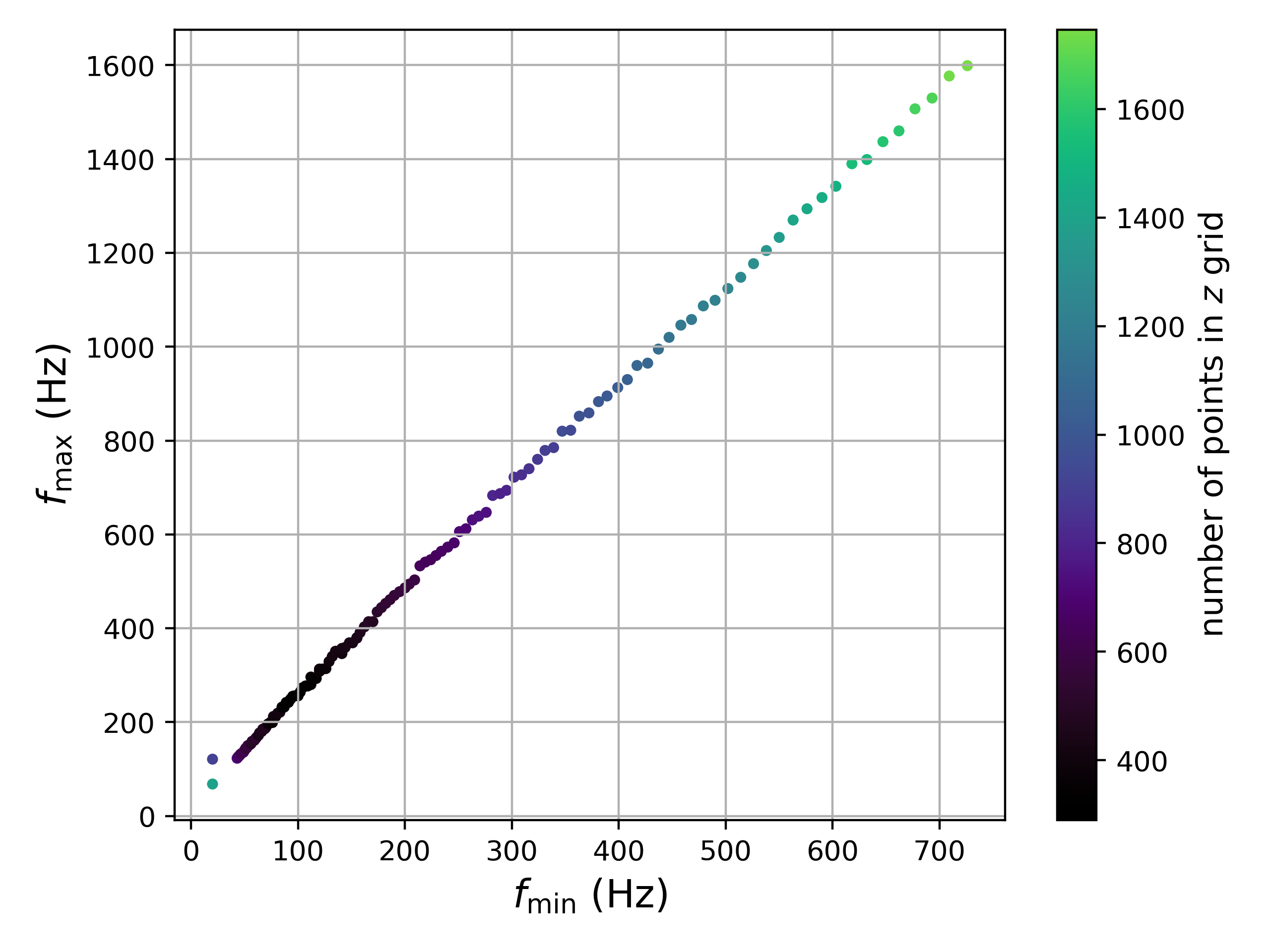}
        }\\ 
    \end{center}
    \caption[]{Left: the number of points in the $k$ grid as a function of the minimum and maximum chirp mass considered, in each configuration. Here, we artificially limit the maximum chirp mass to be $10^{-2}\msun$, resulting in a concentration of points for large $\Mc_{\rm min}$. Furthermore, the one point above the others results because the corresponding $\TFFT$ for that chirp mass range is 29 seconds, compared to the others which are less than 10 seconds. Such a long $\TFFT$ implies more points in the $k$ grid \cite{Miller:2018rbg}. Right: the number of points in the $z$ grid as a function of the frequency range analyzed. 
    }%
     \label{fig:allptsparm}
\end{figure*}

We select candidates above a certain threshold CR, $CR_{\rm thr}\in [7.25,7.67]$, that varies in each configuration to ensure a false alarm probability of 0.1\% in Gaussian noise accounting for the trials factor. $CR_{\rm thr}$ varies in each configuration because the number of points in the parameter space are different, as shown in \cref{fig:allptsparm}, since each configuration probes a different frequency band and chirp mass range \cite{miller_2024_10724845}. 

\begin{figure}
    \centering
    \includegraphics[width=0.49\textwidth]{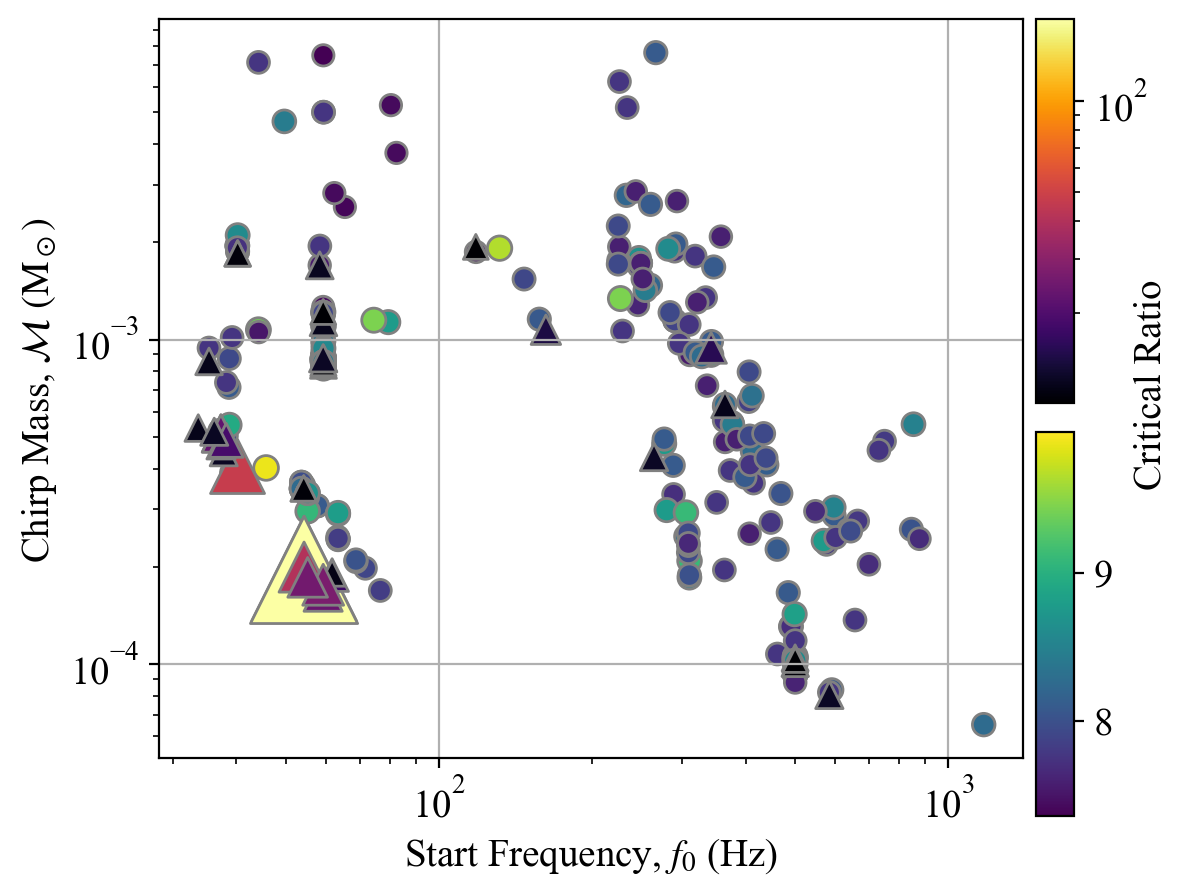}
    \caption{The 334 candidates recovered \cite{miller_2024_10724845}. The color bars indicate the average critical ratio between LLO and LHO, and the two colorbars are split at a $CR$ of 10 (circles for $CR<10$ and triangles for $CR>10$). The size of the markers increases with $CR$.}
    \label{fig:cands2fu}
\end{figure}

After this procedure, 7457 candidates remain. We then veto candidates within a frequency bin of known noise lines \cite{Sun:2020wke}, which leaves us with 334 candidates. We show in \cref{fig:cands2fu} the chirp masses and initial frequencies of these candidates, with their average critical ratios colored, which correspond to average matched-filter \snr of $\rho_{\rm opt}\sim [25,240]$, where

\begin{equation}
    \rho_{\rm opt}^2 = 4\int_{\fmin}^{\fmax} df \frac{|\tilde{h}(f)|^2}{S_n(f)}.
\end{equation}
Here, $\tilde{h}(f)$ is the Fourier transform of the \gwh template $h(t)$.

We now perform a follow-up of these interesting candidates by using the recovered $f_0$ and $\Mc$ from each candidate to demodulate the original time-series data $h(t)$ by multiplying it by a complex exponential that contains the phase evolution $\phi(t)=\phi(t,t_0,f_0,\Mc)$ of the signal whose parameters we have recovered:

\begin{align}
    h^{\rm het}(t) &= \text{Re}[h(t)e^{-i\phi(t)}] \\
    \phi(t)&=2\pi\int_{t_0}^{\Tobs+t_0} (f_{\rm gw}(t)-f_0)dt
\end{align}
where $h^{\rm het}(t)$ is the new time-series that has been demodulated using the expected frequency evolution of the candidate.
This procedure is called heterodyning \cite{Dupuis:2005xv, Miller:2018rbg,LIGOScientific:2018urg,Piccinni:2018akm,LIGOScientific:2020gml,LIGOScientific:2021hvc}. 

We do this demodulation by taking an inverse FFT of the SFDBs to arrive in the time domain, and then applying the heterodyning. If the parameters in $\phi(t)$ exactly match the signal parameters in the data, all the frequency modulation would be removed, resulting in a purely monochromatic signal. At this point, since the signal would be sinusoidal, we could simply take a single FFT and look for a peak at 0 Hz.

In practice, however, we will never be able to perform a perfect correction, since there are uncertainties on the recovered $z_0$ and $k$ for each candidate. Thus, we follow up candidates iteratively by doubling the FFT in each iteration. In the first iteration, $T_{\rm FFT, new}=2\TFFT$, where $\TFFT\in[2,29]$ is the original FFT length used in core stage of the search, which varied in each configuration; in the second iteration, $T_{\rm FFT,new,2}=2T_{\rm FFT, new}$, and so on.

We choose to double $\TFFT$ in each pass because of the uncertainties in the estimated parameters. If we increase $\TFFT$ too much, then that one candidate would correspond to many more that we would need to follow up. As an example, imagine that the only parameter was frequency, and we recover a signal with $f_0$ = 100 Hz $\pm1/\TFFT$ Hz and let $\TFFT = 2$ s. If we then double $\TFFT$, the true signal frequency could be 100, 100.25, 100.5, 99.75 or 99.5 Hz, meaning that we would have to perform 5 follow-ups for one candidate, i.e. we would effectively have to run 5 separate searches for each candidate. We can therefore see how the computing cost can explode if we increase $\TFFT$ too much in each pass. The choice to double $\TFFT$ is certainly an arbitrary one, but we want to keep the number of candidates to follow up under control. Since, in this example, we have already detected this signal with $\TFFT = 2$ s, and if the signal is truly there, the significance will improve with increasing $\TFFT$, as seen in \cref{eqn:dmax}.

\begin{figure}
    \centering
    \includegraphics[width=0.49\textwidth]{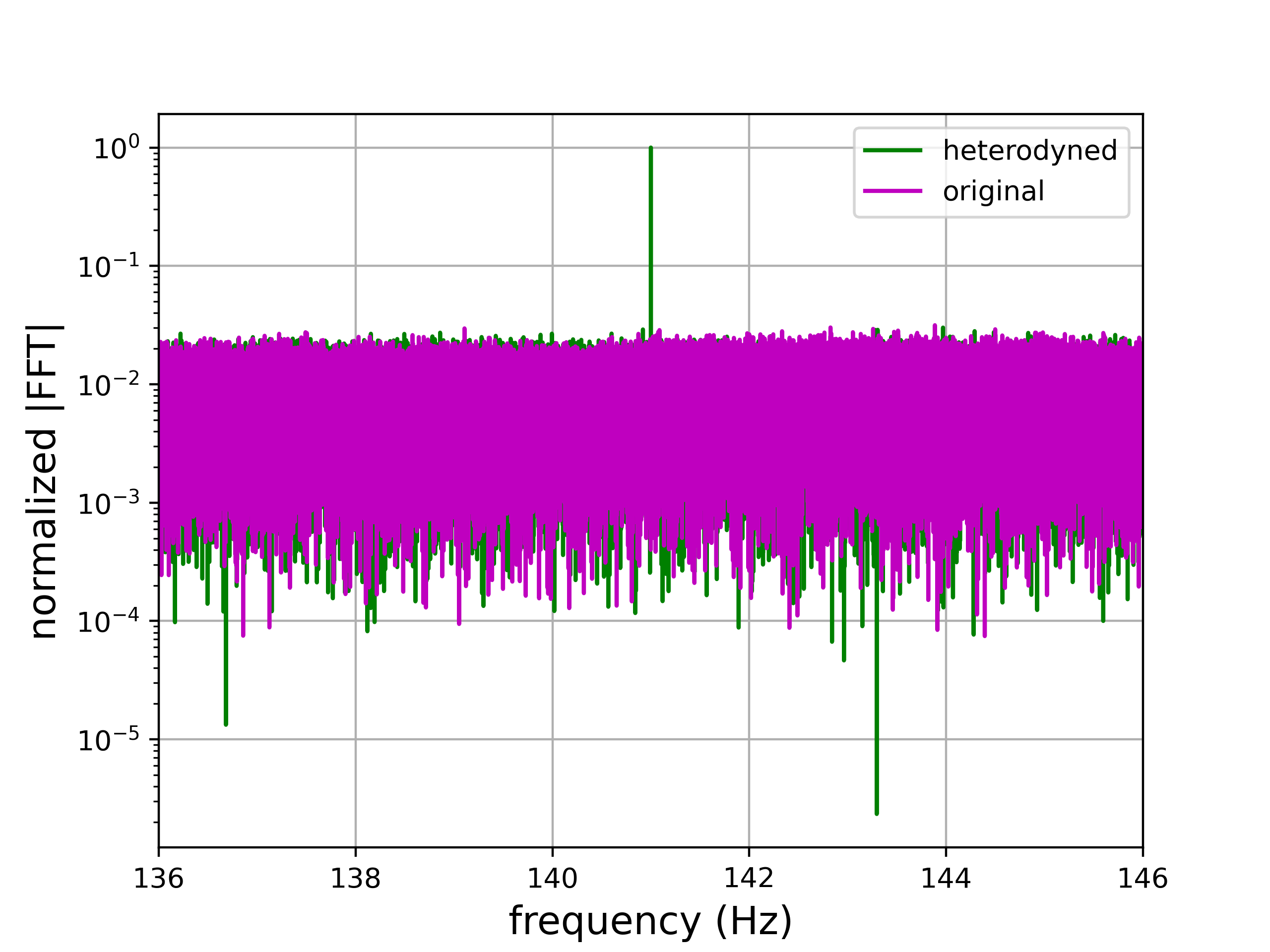}
    \caption{Normalized amplitude spectrum of a simulated inspiraling \pbh siganl before and after heterodyning (demodulating the time series for the expected signal parameters). {We have shifted the frequency axis by $f_0$ so that the reader can see the signal peaked at the \gwh frequency $f_0$.} The normalization is done by dividing each value of the amplitude power spectrum by the maximum value.}
    \label{fig:testhet}
\end{figure}

We show in \cref{fig:testhet} an example of the power spectrum of a heterodyned signal. Before heterodyning, the signal is not visible because $\TFFT\gg1/\sqrt{2\dot{f}}$; however, once heterodyning is performed, the modulation of the signal has been removed, thus making it monochromatic and visible in this power spectrum by eye.

After doubling $\TFFT$, we create another time/frequency peakmap and perform the original \fh transform \cite{Astone:2014esa}. This accounts for some residual spin-up or spin-down in the signal due to some mismatch in the true signal parameters and the ones used for the heterodyning. If the new $CR$ is smaller than the old $CR$, we veto the candidate. In our follow-up procedure, all 334 candidates were vetoed after a single heterodyning.

\section{Upper limits}\label{app:ul}

\begin{figure}
    \centering
    \includegraphics[width=0.49\textwidth]{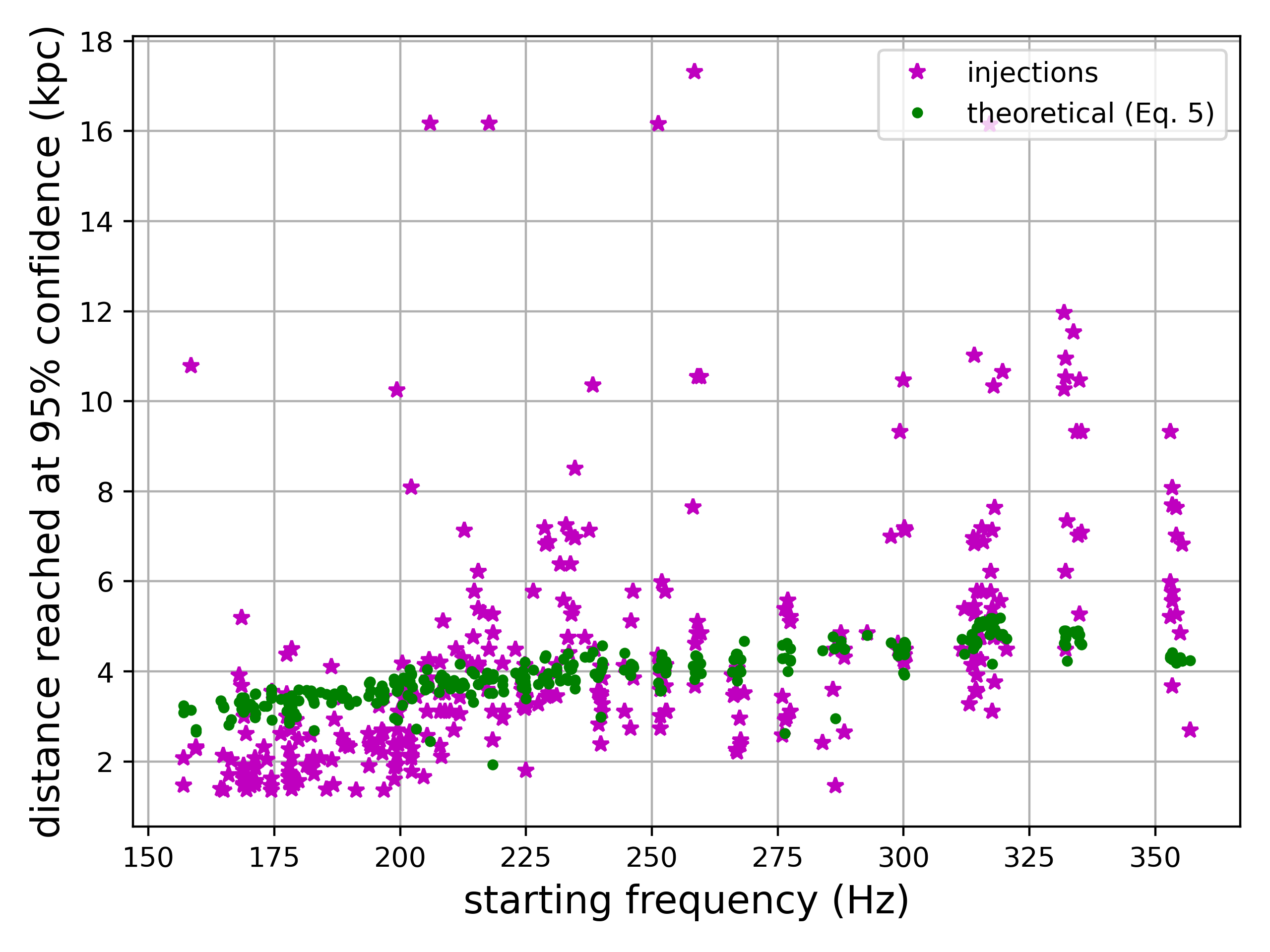}
    \caption{Comparison of injections to our theoretical upper limit procedure for $\Mc=10^{-3}M_\odot$, source duration of 5300 seconds, $\TFFT=2$ s, and a frequency range of [155, 379] Hz. The error on the distance reach of each injection are around $10\%$ of the injected distance reach, which comes from the spacing in signal amplitude that was used to do the simulations.}
    \label{fig:d95_theo_inj}
\end{figure}

In \cite{Miller:2024fpo}, we claim, through an analytic procedure, that our constraints represent 95\% confidence-level upper limits on the distance away from us that we could have detected \gws from inspiraling \pbh systems. In order to verify this claim, we perform injections based on the candidate parameters found in one of our configurations, at a fixed chirp mass of $10^{-3}M_\odot$, across the six months of O3a data. 
For each candidate that we recover in the real search in this configuration and for a range of distances centered around the one derived in our analytic upper limit procedure, we perform 50 injections in LHO data. We require that we recover the candidate within 3 bins of the injection (see \cref{eqn:dbin}), and that the CR of the injection is greater than the CR of the candidate returned in the search. We find generally good agreement between the theoretical and empirical approaches: after accounting for the non-science data and duty cycle ($\sim70\%$) of the detector, and the fact LHO is about a factor of $\sim 1.5$ less sensitive in strain with respect to LLO \cite{LIGOScientific:2020ibl}, we find a median distance reach of $3.67$ kpc with standard deviation of $2.7$ kpc using injections, and a median of $3.85$ kpc with standard deviation of $0.59$ kpc from our upper limit procedure. 
In \cref{fig:d95_theo_inj}, we show, candidate by candidate, the distance reach at 95\% confidence obtained theoretically through our procedure and with injections. 

The upper limits represent a sort of average over all possible signals that could exist in the peakmap over all times. This is because the peakmap is constructed to be optimally sensitive to particular chirp mass systems but is actually sensitive to a range of them, depending on which frequencies the signal spans during $\Tobs$. When doing injections, we have randomly selected different $f_0$ for the one $\Mc$, at different times, meaning that the noise PSD varies, that different injections last for different durations in the peakmap, and that different injections will have different amplitudes. That is why the injections lie on either side of the theoretical curve.

As we can see, for the same chirp mass system, choosing different starting frequencies and times to inject the signal significantly impacts the distance reached. We choose the median as an approximate for the true sensitivity to any signal present in a given peakmap, and note that since the peakmaps overlap in time and in frequency, that is, they are optimized for particular chirp masses, the median actually represents a conservative upper limit with respect to that which would be obtained if we used the upper limit calculated exactly for the kinds of signals for which the peakmap was optimized to detect, which are essentially the points in \cref{fig:d95_theo_inj} with the highest distance reach. 

Additionally, when computing the upper limits analytically in \cite{Miller:2024fpo}, we obtain a median distance reach for each configuration, for each chirp mass. We then take the maximum of the median distance reach of each configuration, which is a fair choice because there is repetition of chirp mass systems we are sensitive to in each configuration -- see the table in \cite{miller_2024_10724845}. Since the search is run in every configuration across all times, we can be confident that we would have detected a signal at a certain distance away from us equal to the maximum of the median distance values in at least one of the configurations.

\begin{figure}
    \centering
    \includegraphics[width=0.49\textwidth]{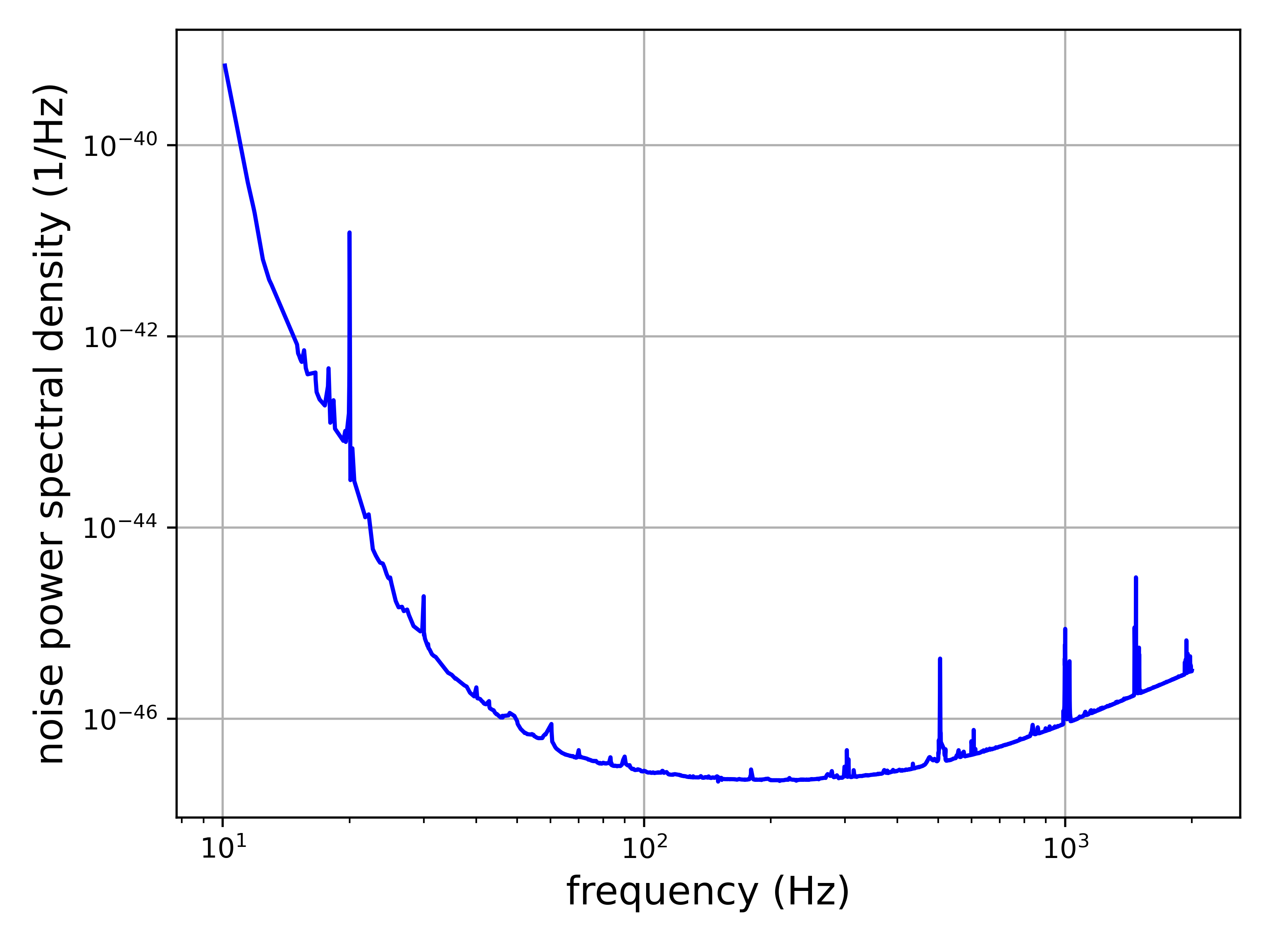}
    \caption{{Noise \psd $S_n(f_{\rm gw})$ used in the upper limit calculations. This PSD is computed by averaging the PSDs computed over all times in LIGO Hanford and LIGO Livingston data, resulting in two effective PSDs, one per detector. These two average PSDs are then averaged together to create this plot.}}
    \label{fig:sn}
\end{figure}

We also provide a plot of the noise \psd that we use to create the upper limit plots in \cite{Miller:2024fpo} in \cref{fig:sn}, which is the average of the LHO/LLO PSDs over O3a.

\section{Computational cost}\label{sec:comp}

The search for \gws from inspiraling \pbhs is computationally inexpensive compared to traditional matched-filtering searches for \ssm binaries. There are a couple of reasons for this: (1) we do not have to create large template banks that contain exactly the waveforms we search for, and (2) we can actually search for multiple waveforms simultaneously on each peakmap with the \GFH, since the \GFH is a semi-coherent method that is only sensitive to time-frequency variations, as opposed to matched filtering, which is extremely sensitive to phase evolutions. Furthermore, the search is also computationally light compared to \cwh searches, since we do not require analyzing each sky point individually, and in fact are sensitive to sources coming from \emph{anywhere} in the sky. The trade-off here is that we cannot localize any sources in the sky, since all Doppler modulations induced by the relative motion of the earth and the source at a particular sky position do not cause the \pbh signal frequency to vary by more than one frequency bin. 

Thus, these comparisons indicate that our search should be much faster than the others, and in fact it is: using 1935 CPUs, our search for planetary-mass \pbhs was completed in $\sim 3-5$ days of wall time (not core-hours).  

\begin{figure}
    \centering
    \includegraphics[width=0.49\textwidth]{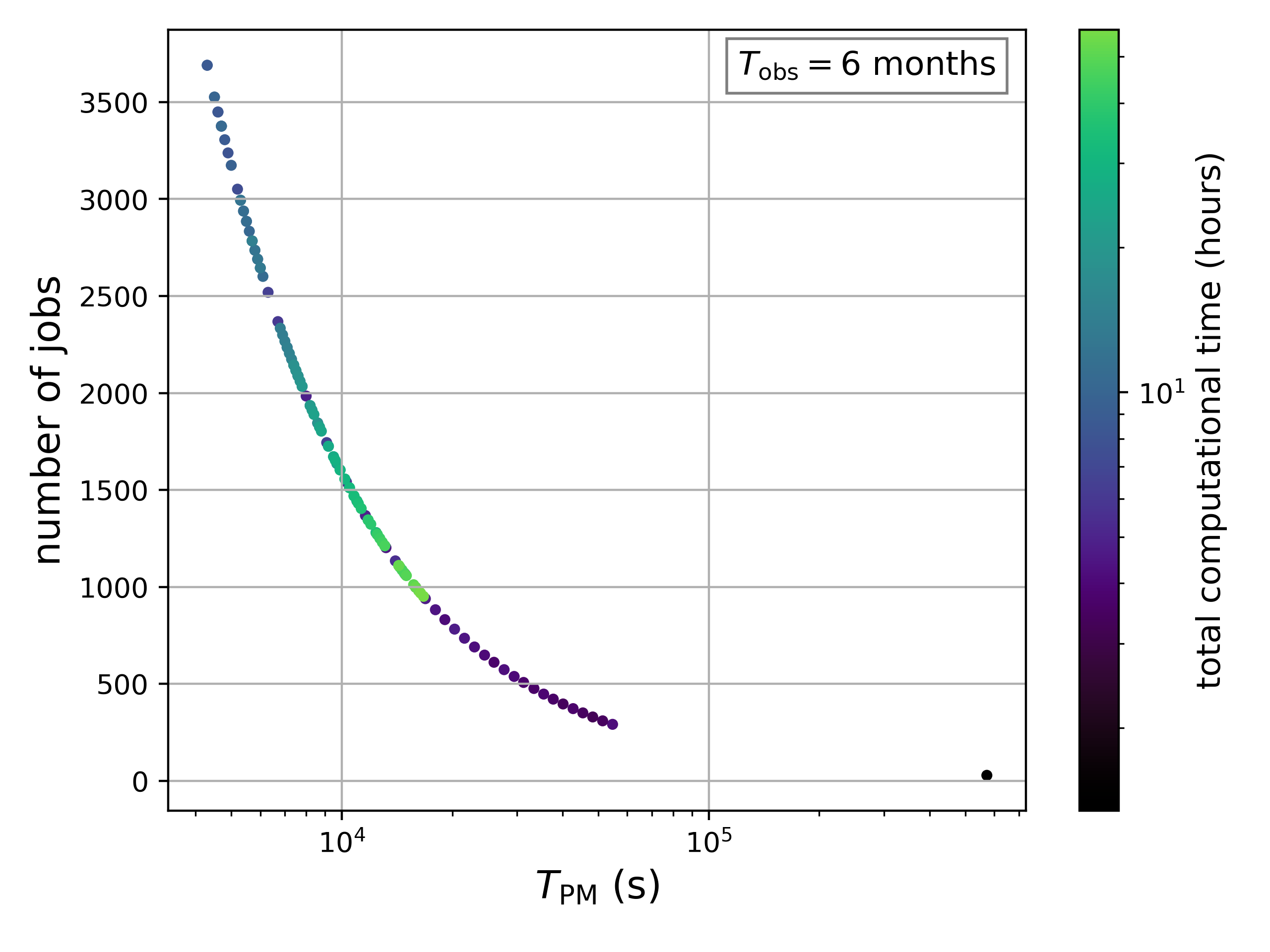}
    \caption{The computational cost to run our search in each configuration is shown as a function of the duration of each peakmap $\Tobs$. The number of jobs that each configuration requires is also plotted as a function of $\Tobs$. Roughly, the total computational cost of the search was 2400 core-hours, and took a few days on 1935 CPUs. }
    \label{fig:compcost}
\end{figure}

To illustrate quantitatively why the computational cost is much lower for this search than for others, we plot in \cref{fig:allptsparm} the number of $k$ and $z$ points in the parameter space in each configuration, in one peakmap. The product of the two color bars represents the total number of chirp masses and frequencies that the \GFH searches over, in each configuration, in each peakmap. Furthermore, each particular point in the parameter space takes about 50$\mu$s to be searched for with the \GFH. We can therefore plot the approximate computational cost in each configuration, and the number of jobs that must be submitted per configuration, as a function of $\Tobs$, which is shown in \cref{fig:compcost}. The calculated computational cost from \cref{fig:compcost}, obtained by summing the color values at each point, is approximately 2400 core-hours. This can be contrasted with $10^7$ core-hours
for the \fh all-sky searches\footnote{We note, however, that deep learning methods could also be used to quicken \cwh searches by orders of magnitude \cite{Dreissigacker:2019edy,Bayley:2022hkz,Yamamoto:2022adl}.} \cite{KAGRA:2022dwb}, thousands of years on a GPU for deeper \cwh searches \cite{Steltner:2023cfk}, and $\sim 900$ million core-hours in an O3a search for \ssm \pbhs between [0.1,1]$\msun$ \cite{LIGOScientific:2021job}. We show the comparison of the computational costs of these different searches in \cref{tab:compcost}.

We will now explain where our estimates of the computational costs of the \fh and \ssm searches come from.  The \fh search only analyzed data up to 1024 Hz in \cite{LIGOScientific:2019yhl}, and required 9 million core-hours. To obtain an estimate in the O3 \fh search in \cite{KAGRA:2022dwb}, we note that the major computational cost of the \fh search arises from the number of points in the sky grid. Using Eq. 43 in \cite{Astone:2014esa} and the corresponding $\TFFT$ lengths given in Tab. I of \cite{LIGOScientific:2019yhl}, we calculate the number of sky points searched over in the O2 search to be $\sim 1.7\times 10^8$. If the \fh had analyzed frequencies up to 2048 Hz in O2, then the number of additional sky points would have been $\sim 3 \times 10^8$. Therefore, we can roughly estimate that the computational cost in core-hours would have been $\sim 1.7$ times larger than the 9 million core-hours given in \cite{LIGOScientific:2019yhl}. There is an additional contribution to the computational cost for the increased size of the grid in $\dot{f}$ in \cite{KAGRA:2022dwb} compared to that in \cite{LIGOScientific:2019yhl}, but that is expected to contribute less to the computational cost of the search than the increase in sky points \cite{Astone:2014esa}.

For the matched-filtering \ssm search, we have made a rough estimate of the computational cost by noting that \cite{LIGOScientific:2019kan} used $7.8\times 10^{6}$ templates, and \cite{LIGOScientific:2021job} used double that number for a six-month search. We calculate that the Fourier transform and matched filtering operation takes $\sim 1.3$ seconds on an Apple M1 CPU for a waveform starting at 45 Hz up to 2048 Hz sampled at 4096 Hz, and assume that each waveform lasts 100 seconds.

\begin{table*}[t]
    \centering
    \caption{Comparison of computational costs for \gwh searches that probe vastly different \ssm regimes. The \GFH is the quickest of all these searches. For the two all-sky searches, we compute the chirp-mass ranges by using the maximum spin-up to which they are sensitive, and requiring that the \gwh frequency evolution given in \cref{eqn:powlaws} does not differ by more than one frequency bin from that in \cref{eqn:linf}.}
    \label{tab:compcost}
    \begin{ruledtabular}
    \begin{tabular}{lll}
        Method & Computational cost (core-hours) & chirp-mass range $(\msun)$ \\ \hline
        \fh O3 all-sky search & $10^7$ \cite{LIGOScientific:2019yhl,KAGRA:2022dwb} & $[1.2\times 10^{-9},3\times 10^{-5}]$ \\
        Einstein@Home O3 all-sky search & $8\times 10^6$ on a GPU \cite{Steltner:2023cfk} & $[4.2\times 10^{-9},1.5\times 10^{-6}]$ \\
        Matched filtering O3a \ssm search & $900\times 10^{6}$ \cite{LIGOScientific:2021job} & $[10^{-1},1]$\\
        \GFH planetary-mass O3a \pbh search & $2.4\times 10^3$ \cite{Miller:2024fpo}& $[4\times 10^{-5},10^{-2}]$ 
    \end{tabular}
    \end{ruledtabular}
\end{table*}

\section{Conclusions}

We have presented an application of the \GFH to search for planetary-mass \pbh systems inspiraling towards one another over time. This companion paper describes our O3a search from start to finish: it explains the method we have used, the \GFH, our search for planetary-mass \pbhs in O3a data, and our techniques to rule out potential candidates of \gwh emission arising from \pbh inspirals. Having found no significant candidates, we also describe how we analytically calculated the upper limits on distance reached presented in \cite{Miller:2024fpo}, with a comparison to empirical estimates of the upper limits through injections. We have shown that computing upper limits through our our analytic procedure produces consistent upper limits with respect to what we would have obtained with injections, and conservative upper limits with respect to the limits we would obtained if we were to inject the optimal \pbh signal to which each peakmap was designed to be sensitive. Finally, we have shown that our search is much less computationally intensive than both \cwh and \ssm searches.



Future work includes applying the \GFH on data from subsequent observing runs of LIGO, Virgo and KAGRA, and expanding the search to include sky position, especially for the systems with $\Mc\sim 10^{-5}\msun$. At the moment, we cannot resolve sky location, since our $\TFFT$ is small enough to confine the frequency modulation due to the Doppler shift of the earth motion with respect to the source location, to one frquency bin for the entire observing run. By including sky position, we can ``point'' to particular areas in which we expect high concentrations of \pbhs, such as globular clusters, or to astrophysical black holes or neutron stars that could have a very light companion orbiting around it. Thus, it is necessary to not only include the Doppler shift in our search, but also to consider searching specifically for highly asymmetric mass ratio systems \cite{Guo:2022sdd}, for which we currently have some sensitivity, though we are not optimally sensitive to them. The inclusion of asymmetric mass ratio systems to specifically search for could greatly improve the constraints presented in \cite{Miller:2024fpo}, since the formation rates for such systems are expected to be orders of magnitude higher than equal-mass systems \cite{Carr:2019kxo}. Furthermore, they may be applications of such \cwh methods to constrain astrophysical objects, e.g. stars, planets, that transmute into \pbhs \cite{Takhistov:2017bpt,Takhistov:2020vxs,Bhattacharya:2024pmp}, using future ground- and space-based detector data.

\section*{Acknowledgments}
This material is based upon work supported by NSF's LIGO Laboratory which is a major facility fully funded by the National Science Foundation

We would like to thank the Rome Virgo group for the tools necessary to perform these studies, such as the development of the original \fh transform and the development of the short FFT databases. Additionally we would like to thank Luca Rei for managing data transfers.

This research has made use of data, software and/or web tools obtained from the Gravitational Wave Open Science Center (https://www.gw-openscience.org/ ), a service of LIGO Laboratory, the LIGO Scientific Collaboration and the Virgo Collaboration. LIGO Laboratory and Advanced LIGO are funded by the United States National Science Foundation (NSF) as well as the Science and Technology Facilities Council (STFC) of the United Kingdom, the Max-Planck-Society (MPS), and the State of Niedersachsen/Germany for support of the construction of Advanced LIGO and construction and operation of the GEO600 detector. Additional support for Advanced LIGO was provided by the Australian Research Council. Virgo is funded, through the European Gravitational Observatory (EGO), by the French Centre National de Recherche Scientifique (CNRS), the Italian Istituto Nazionale della Fisica Nucleare (INFN) and the Dutch Nikhef, with contributions by institutions from Belgium, Germany, Greece, Hungary, Ireland, Japan, Monaco, Poland, Portugal, Spain.

We also wish to acknowledge the support of the INFN-CNAF computing center for its help with the storage and transfer of the data used in this paper.

We would like to thank all of the essential workers who put their health at risk during the COVID-19 pandemic, without whom we would not have been able to complete this work.

F.D.L. is supported by a FRIA (Fonds pour la formation à la Recherche dans l'Industrie et dans l'Agriculture) Grant of the Belgian Fund for Research, F.R.S.-FNRS (Fonds de la Recherche Scientifique-FNRS).

This work is partially supported by ICSC – Centro Nazionale di Ricerca in High Performance Computing, Big Data and Quantum Computing, funded by European Union – NextGenerationEU.

\bibliographystyle{apsrev4-1}
\bibliography{biblio,biblio_pbh_method}

\begin{thebibliography}{72}%
\makeatletter
\providecommand \@ifxundefined [1]{%
 \@ifx{#1\undefined}
}%
\providecommand \@ifnum [1]{%
 \ifnum #1\expandafter \@firstoftwo
 \else \expandafter \@secondoftwo
 \fi
}%
\providecommand \@ifx [1]{%
 \ifx #1\expandafter \@firstoftwo
 \else \expandafter \@secondoftwo
 \fi
}%
\providecommand \natexlab [1]{#1}%
\providecommand \enquote  [1]{``#1''}%
\providecommand \bibnamefont  [1]{#1}%
\providecommand \bibfnamefont [1]{#1}%
\providecommand \citenamefont [1]{#1}%
\providecommand \href@noop [0]{\@secondoftwo}%
\providecommand \href [0]{\begingroup \@sanitize@url \@href}%
\providecommand \@href[1]{\@@startlink{#1}\@@href}%
\providecommand \@@href[1]{\endgroup#1\@@endlink}%
\providecommand \@sanitize@url [0]{\catcode `\\12\catcode `\$12\catcode `\&12\catcode `\#12\catcode `\^12\catcode `\_12\catcode `\%12\relax}%
\providecommand \@@startlink[1]{}%
\providecommand \@@endlink[0]{}%
\providecommand \url  [0]{\begingroup\@sanitize@url \@url }%
\providecommand \@url [1]{\endgroup\@href {#1}{\urlprefix }}%
\providecommand \urlprefix  [0]{URL }%
\providecommand \Eprint [0]{\href }%
\providecommand \doibase [0]{http://dx.doi.org/}%
\providecommand \selectlanguage [0]{\@gobble}%
\providecommand \bibinfo  [0]{\@secondoftwo}%
\providecommand \bibfield  [0]{\@secondoftwo}%
\providecommand \translation [1]{[#1]}%
\providecommand \BibitemOpen [0]{}%
\providecommand \bibitemStop [0]{}%
\providecommand \bibitemNoStop [0]{.\EOS\space}%
\providecommand \EOS [0]{\spacefactor3000\relax}%
\providecommand \BibitemShut  [1]{\csname bibitem#1\endcsname}%
\let\auto@bib@innerbib\@empty
\bibitem [{\citenamefont {Miller}\ \emph {et~al.}(2024{\natexlab{a}})\citenamefont {Miller}, \citenamefont {Aggarwal}, \citenamefont {Clesse}, \citenamefont {De~Lillo}, \citenamefont {Sachdev}, \citenamefont {Astone}, \citenamefont {Palomba}, \citenamefont {Piccinni},\ and\ \citenamefont {Pierini}}]{Miller:2024fpo}%
  \BibitemOpen
  \bibfield  {author} {\bibinfo {author} {\bibfnamefont {A.~L.}\ \bibnamefont {Miller}}, \bibinfo {author} {\bibfnamefont {N.}~\bibnamefont {Aggarwal}}, \bibinfo {author} {\bibfnamefont {S.}~\bibnamefont {Clesse}}, \bibinfo {author} {\bibfnamefont {F.}~\bibnamefont {De~Lillo}}, \bibinfo {author} {\bibfnamefont {S.}~\bibnamefont {Sachdev}}, \bibinfo {author} {\bibfnamefont {P.}~\bibnamefont {Astone}}, \bibinfo {author} {\bibfnamefont {C.}~\bibnamefont {Palomba}}, \bibinfo {author} {\bibfnamefont {O.~J.}\ \bibnamefont {Piccinni}}, \ and\ \bibinfo {author} {\bibfnamefont {L.}~\bibnamefont {Pierini}},\ }\href@noop {} {\  (\bibinfo {year} {2024}{\natexlab{a}})},\ \Eprint {http://arxiv.org/abs/2402.19468} {arXiv:2402.19468 [gr-qc]} \BibitemShut {NoStop}%
\bibitem [{\citenamefont {Bertone}\ \emph {et~al.}(2005)\citenamefont {Bertone}, \citenamefont {Hooper},\ and\ \citenamefont {Silk}}]{BERTONE2005279}%
  \BibitemOpen
  \bibfield  {author} {\bibinfo {author} {\bibfnamefont {G.}~\bibnamefont {Bertone}}, \bibinfo {author} {\bibfnamefont {D.}~\bibnamefont {Hooper}}, \ and\ \bibinfo {author} {\bibfnamefont {J.}~\bibnamefont {Silk}},\ }\href {\doibase https://doi.org/10.1016/j.physrep.2004.08.031} {\bibfield  {journal} {\bibinfo  {journal} {Physics Reports}\ }\textbf {\bibinfo {volume} {405}},\ \bibinfo {pages} {279} (\bibinfo {year} {2005})}\BibitemShut {NoStop}%
\bibitem [{\citenamefont {Bertone}\ and\ \citenamefont {Tait}(2018)}]{bertone2018new}%
  \BibitemOpen
  \bibfield  {author} {\bibinfo {author} {\bibfnamefont {G.}~\bibnamefont {Bertone}}\ and\ \bibinfo {author} {\bibfnamefont {T.~M.}\ \bibnamefont {Tait}},\ }\href@noop {} {\bibfield  {journal} {\bibinfo  {journal} {Nature}\ }\textbf {\bibinfo {volume} {562}},\ \bibinfo {pages} {51} (\bibinfo {year} {2018})}\BibitemShut {NoStop}%
\bibitem [{\citenamefont {Klasen}\ \emph {et~al.}(2015)\citenamefont {Klasen}, \citenamefont {Pohl},\ and\ \citenamefont {Sigl}}]{KLASEN20151}%
  \BibitemOpen
  \bibfield  {author} {\bibinfo {author} {\bibfnamefont {M.}~\bibnamefont {Klasen}}, \bibinfo {author} {\bibfnamefont {M.}~\bibnamefont {Pohl}}, \ and\ \bibinfo {author} {\bibfnamefont {G.}~\bibnamefont {Sigl}},\ }\href {\doibase https://doi.org/10.1016/j.ppnp.2015.07.001} {\bibfield  {journal} {\bibinfo  {journal} {Progress in Particle and Nuclear Physics}\ }\textbf {\bibinfo {volume} {85}},\ \bibinfo {pages} {1} (\bibinfo {year} {2015})}\BibitemShut {NoStop}%
\bibitem [{\citenamefont {Carr}\ \emph {et~al.}(2016)\citenamefont {Carr}, \citenamefont {Kuhnel},\ and\ \citenamefont {Sandstad}}]{Carr:2016drx}%
  \BibitemOpen
  \bibfield  {author} {\bibinfo {author} {\bibfnamefont {B.}~\bibnamefont {Carr}}, \bibinfo {author} {\bibfnamefont {F.}~\bibnamefont {Kuhnel}}, \ and\ \bibinfo {author} {\bibfnamefont {M.}~\bibnamefont {Sandstad}},\ }\href {\doibase 10.1103/PhysRevD.94.083504} {\bibfield  {journal} {\bibinfo  {journal} {Phys. Rev. D}\ }\textbf {\bibinfo {volume} {94}},\ \bibinfo {pages} {083504} (\bibinfo {year} {2016})},\ \Eprint {http://arxiv.org/abs/1607.06077} {arXiv:1607.06077 [astro-ph.CO]} \BibitemShut {NoStop}%
\bibitem [{\citenamefont {Bagui}\ \emph {et~al.}(2023)\citenamefont {Bagui} \emph {et~al.}}]{LISACosmologyWorkingGroup:2023njw}%
  \BibitemOpen
  \bibfield  {author} {\bibinfo {author} {\bibfnamefont {E.}~\bibnamefont {Bagui}} \emph {et~al.} (\bibinfo {collaboration} {LISA Cosmology Working Group}),\ }\href@noop {} {\  (\bibinfo {year} {2023})},\ \Eprint {http://arxiv.org/abs/2310.19857} {arXiv:2310.19857 [astro-ph.CO]} \BibitemShut {NoStop}%
\bibitem [{\citenamefont {Yamamoto}\ \emph {et~al.}(2023)\citenamefont {Yamamoto}, \citenamefont {Inui}, \citenamefont {Tada},\ and\ \citenamefont {Yokoyama}}]{Yamamoto:2023tsr}%
  \BibitemOpen
  \bibfield  {author} {\bibinfo {author} {\bibfnamefont {T.~S.}\ \bibnamefont {Yamamoto}}, \bibinfo {author} {\bibfnamefont {R.}~\bibnamefont {Inui}}, \bibinfo {author} {\bibfnamefont {Y.}~\bibnamefont {Tada}}, \ and\ \bibinfo {author} {\bibfnamefont {S.}~\bibnamefont {Yokoyama}},\ }\href@noop {} {\  (\bibinfo {year} {2023})},\ \Eprint {http://arxiv.org/abs/2401.00044} {arXiv:2401.00044 [gr-qc]} \BibitemShut {NoStop}%
\bibitem [{\citenamefont {Carr}(1981)}]{carr:1981xxx}%
  \BibitemOpen
  \bibfield  {author} {\bibinfo {author} {\bibfnamefont {B.~J.}\ \bibnamefont {Carr}},\ }\href {\doibase 10.1093/mnras/194.3.639} {\bibfield  {journal} {\bibinfo  {journal} {Monthly Notices of the Royal Astronomical Society}\ }\textbf {\bibinfo {volume} {194}},\ \bibinfo {pages} {639} (\bibinfo {year} {1981})},\ \Eprint {http://arxiv.org/abs/https://academic.oup.com/mnras/article-pdf/194/3/639/3214058/mnras194-0639.pdf} {https://academic.oup.com/mnras/article-pdf/194/3/639/3214058/mnras194-0639.pdf} \BibitemShut {NoStop}%
\bibitem [{\citenamefont {Ricotti}\ \emph {et~al.}(2008)\citenamefont {Ricotti}, \citenamefont {Ostriker},\ and\ \citenamefont {Mack}}]{Ricotti:2007au}%
  \BibitemOpen
  \bibfield  {author} {\bibinfo {author} {\bibfnamefont {M.}~\bibnamefont {Ricotti}}, \bibinfo {author} {\bibfnamefont {J.~P.}\ \bibnamefont {Ostriker}}, \ and\ \bibinfo {author} {\bibfnamefont {K.~J.}\ \bibnamefont {Mack}},\ }\href {\doibase 10.1086/587831} {\bibfield  {journal} {\bibinfo  {journal} {Astrophys. J.}\ }\textbf {\bibinfo {volume} {680}},\ \bibinfo {pages} {829} (\bibinfo {year} {2008})},\ \Eprint {http://arxiv.org/abs/0709.0524} {arXiv:0709.0524 [astro-ph]} \BibitemShut {NoStop}%
\bibitem [{\citenamefont {Gaggero}\ \emph {et~al.}(2017)\citenamefont {Gaggero}, \citenamefont {Bertone}, \citenamefont {Calore}, \citenamefont {Connors}, \citenamefont {Lovell}, \citenamefont {Markoff},\ and\ \citenamefont {Storm}}]{Gaggero:2016dpq}%
  \BibitemOpen
  \bibfield  {author} {\bibinfo {author} {\bibfnamefont {D.}~\bibnamefont {Gaggero}}, \bibinfo {author} {\bibfnamefont {G.}~\bibnamefont {Bertone}}, \bibinfo {author} {\bibfnamefont {F.}~\bibnamefont {Calore}}, \bibinfo {author} {\bibfnamefont {R.~M.~T.}\ \bibnamefont {Connors}}, \bibinfo {author} {\bibfnamefont {M.}~\bibnamefont {Lovell}}, \bibinfo {author} {\bibfnamefont {S.}~\bibnamefont {Markoff}}, \ and\ \bibinfo {author} {\bibfnamefont {E.}~\bibnamefont {Storm}},\ }\href {\doibase 10.1103/PhysRevLett.118.241101} {\bibfield  {journal} {\bibinfo  {journal} {Phys. Rev. Lett.}\ }\textbf {\bibinfo {volume} {118}},\ \bibinfo {pages} {241101} (\bibinfo {year} {2017})},\ \Eprint {http://arxiv.org/abs/1612.00457} {arXiv:1612.00457 [astro-ph.HE]} \BibitemShut {NoStop}%
\bibitem [{\citenamefont {Page}\ and\ \citenamefont {Hawking}(1976)}]{Page:1976wx}%
  \BibitemOpen
  \bibfield  {author} {\bibinfo {author} {\bibfnamefont {D.~N.}\ \bibnamefont {Page}}\ and\ \bibinfo {author} {\bibfnamefont {S.~W.}\ \bibnamefont {Hawking}},\ }\href {\doibase 10.1086/154350} {\bibfield  {journal} {\bibinfo  {journal} {Astrophys. J.}\ }\textbf {\bibinfo {volume} {206}},\ \bibinfo {pages} {1} (\bibinfo {year} {1976})}\BibitemShut {NoStop}%
\bibitem [{\citenamefont {Boudaud}\ and\ \citenamefont {Cirelli}(2019)}]{Boudaud:2018hqb}%
  \BibitemOpen
  \bibfield  {author} {\bibinfo {author} {\bibfnamefont {M.}~\bibnamefont {Boudaud}}\ and\ \bibinfo {author} {\bibfnamefont {M.}~\bibnamefont {Cirelli}},\ }\href {\doibase 10.1103/PhysRevLett.122.041104} {\bibfield  {journal} {\bibinfo  {journal} {Phys. Rev. Lett.}\ }\textbf {\bibinfo {volume} {122}},\ \bibinfo {pages} {041104} (\bibinfo {year} {2019})},\ \Eprint {http://arxiv.org/abs/1807.03075} {arXiv:1807.03075 [astro-ph.HE]} \BibitemShut {NoStop}%
\bibitem [{\citenamefont {Brandt}(2016)}]{Brandt:2016aco}%
  \BibitemOpen
  \bibfield  {author} {\bibinfo {author} {\bibfnamefont {T.~D.}\ \bibnamefont {Brandt}},\ }\href {\doibase 10.3847/2041-8205/824/2/L31} {\bibfield  {journal} {\bibinfo  {journal} {Astrophys. J. Lett.}\ }\textbf {\bibinfo {volume} {824}},\ \bibinfo {pages} {L31} (\bibinfo {year} {2016})},\ \Eprint {http://arxiv.org/abs/1605.03665} {arXiv:1605.03665 [astro-ph.GA]} \BibitemShut {NoStop}%
\bibitem [{\citenamefont {Koushiappas}\ and\ \citenamefont {Loeb}(2017)}]{Koushiappas:2017chw}%
  \BibitemOpen
  \bibfield  {author} {\bibinfo {author} {\bibfnamefont {S.~M.}\ \bibnamefont {Koushiappas}}\ and\ \bibinfo {author} {\bibfnamefont {A.}~\bibnamefont {Loeb}},\ }\href {\doibase 10.1103/PhysRevLett.119.041102} {\bibfield  {journal} {\bibinfo  {journal} {Phys. Rev. Lett.}\ }\textbf {\bibinfo {volume} {119}},\ \bibinfo {pages} {041102} (\bibinfo {year} {2017})},\ \Eprint {http://arxiv.org/abs/1704.01668} {arXiv:1704.01668 [astro-ph.GA]} \BibitemShut {NoStop}%
\bibitem [{\citenamefont {Tisserand}\ \emph {et~al.}(2007)\citenamefont {Tisserand}, \citenamefont {Le~Guillou}, \citenamefont {Afonso}, \citenamefont {Albert}, \citenamefont {Andersen}, \citenamefont {Ansari}, \citenamefont {Aubourg}, \citenamefont {Bareyre}, \citenamefont {Beaulieu}, \citenamefont {Charlot} \emph {et~al.}}]{tisserand2007limits}%
  \BibitemOpen
  \bibfield  {author} {\bibinfo {author} {\bibfnamefont {P.}~\bibnamefont {Tisserand}}, \bibinfo {author} {\bibfnamefont {L.}~\bibnamefont {Le~Guillou}}, \bibinfo {author} {\bibfnamefont {C.}~\bibnamefont {Afonso}}, \bibinfo {author} {\bibfnamefont {J.}~\bibnamefont {Albert}}, \bibinfo {author} {\bibfnamefont {J.}~\bibnamefont {Andersen}}, \bibinfo {author} {\bibfnamefont {R.}~\bibnamefont {Ansari}}, \bibinfo {author} {\bibfnamefont {{\'E}.}~\bibnamefont {Aubourg}}, \bibinfo {author} {\bibfnamefont {P.}~\bibnamefont {Bareyre}}, \bibinfo {author} {\bibfnamefont {J.}~\bibnamefont {Beaulieu}}, \bibinfo {author} {\bibfnamefont {X.}~\bibnamefont {Charlot}},  \emph {et~al.},\ }\href@noop {} {\bibfield  {journal} {\bibinfo  {journal} {Astronomy \& Astrophysics}\ }\textbf {\bibinfo {volume} {469}},\ \bibinfo {pages} {387} (\bibinfo {year} {2007})}\BibitemShut {NoStop}%
\bibitem [{\citenamefont {Niikura}\ \emph {et~al.}(2019)\citenamefont {Niikura}, \citenamefont {Takada}, \citenamefont {Yokoyama}, \citenamefont {Sumi},\ and\ \citenamefont {Masaki}}]{niikura2019constraints}%
  \BibitemOpen
  \bibfield  {author} {\bibinfo {author} {\bibfnamefont {H.}~\bibnamefont {Niikura}}, \bibinfo {author} {\bibfnamefont {M.}~\bibnamefont {Takada}}, \bibinfo {author} {\bibfnamefont {S.}~\bibnamefont {Yokoyama}}, \bibinfo {author} {\bibfnamefont {T.}~\bibnamefont {Sumi}}, \ and\ \bibinfo {author} {\bibfnamefont {S.}~\bibnamefont {Masaki}},\ }\href@noop {} {\bibfield  {journal} {\bibinfo  {journal} {Physical Review D}\ }\textbf {\bibinfo {volume} {99}},\ \bibinfo {pages} {083503} (\bibinfo {year} {2019})}\BibitemShut {NoStop}%
\bibitem [{\citenamefont {Croon}\ \emph {et~al.}(2020)\citenamefont {Croon}, \citenamefont {McKeen}, \citenamefont {Raj},\ and\ \citenamefont {Wang}}]{Croon:2020ouk}%
  \BibitemOpen
  \bibfield  {author} {\bibinfo {author} {\bibfnamefont {D.}~\bibnamefont {Croon}}, \bibinfo {author} {\bibfnamefont {D.}~\bibnamefont {McKeen}}, \bibinfo {author} {\bibfnamefont {N.}~\bibnamefont {Raj}}, \ and\ \bibinfo {author} {\bibfnamefont {Z.}~\bibnamefont {Wang}},\ }\href {\doibase 10.1103/PhysRevD.102.083021} {\bibfield  {journal} {\bibinfo  {journal} {Phys. Rev. D}\ }\textbf {\bibinfo {volume} {102}},\ \bibinfo {pages} {083021} (\bibinfo {year} {2020})},\ \Eprint {http://arxiv.org/abs/2007.12697} {arXiv:2007.12697 [astro-ph.CO]} \BibitemShut {NoStop}%
\bibitem [{\citenamefont {Arvanitaki}\ \emph {et~al.}(2015)\citenamefont {Arvanitaki}, \citenamefont {Huang},\ and\ \citenamefont {Van~Tilburg}}]{Arvanitaki:2014faa}%
  \BibitemOpen
  \bibfield  {author} {\bibinfo {author} {\bibfnamefont {A.}~\bibnamefont {Arvanitaki}}, \bibinfo {author} {\bibfnamefont {J.}~\bibnamefont {Huang}}, \ and\ \bibinfo {author} {\bibfnamefont {K.}~\bibnamefont {Van~Tilburg}},\ }\href {\doibase 10.1103/PhysRevD.91.015015} {\bibfield  {journal} {\bibinfo  {journal} {Phys. Rev. D}\ }\textbf {\bibinfo {volume} {91}},\ \bibinfo {pages} {015015} (\bibinfo {year} {2015})},\ \Eprint {http://arxiv.org/abs/1405.2925} {arXiv:1405.2925 [hep-ph]} \BibitemShut {NoStop}%
\bibitem [{\citenamefont {Green}\ and\ \citenamefont {Kavanagh}(2021)}]{Green:2020jor}%
  \BibitemOpen
  \bibfield  {author} {\bibinfo {author} {\bibfnamefont {A.~M.}\ \bibnamefont {Green}}\ and\ \bibinfo {author} {\bibfnamefont {B.~J.}\ \bibnamefont {Kavanagh}},\ }\href {\doibase 10.1088/1361-6471/abc534} {\bibfield  {journal} {\bibinfo  {journal} {J. Phys. G}\ }\textbf {\bibinfo {volume} {48}},\ \bibinfo {pages} {043001} (\bibinfo {year} {2021})},\ \Eprint {http://arxiv.org/abs/2007.10722} {arXiv:2007.10722 [astro-ph.CO]} \BibitemShut {NoStop}%
\bibitem [{\citenamefont {Abbott}\ \emph {et~al.}(2016)\citenamefont {Abbott} \emph {et~al.}}]{Abbott:2016blz}%
  \BibitemOpen
  \bibfield  {author} {\bibinfo {author} {\bibfnamefont {B.}~\bibnamefont {Abbott}} \emph {et~al.} (\bibinfo {collaboration} {LIGO Scientific Collaboration, Virgo}),\ }\href {\doibase 10.1103/PhysRevLett.116.061102} {\bibfield  {journal} {\bibinfo  {journal} {Phys. Rev. Lett.}\ }\textbf {\bibinfo {volume} {116}},\ \bibinfo {pages} {061102} (\bibinfo {year} {2016})},\ \Eprint {http://arxiv.org/abs/1602.03837} {arXiv:1602.03837 [gr-qc]} \BibitemShut {NoStop}%
\bibitem [{\citenamefont {Sasaki}\ \emph {et~al.}(2016)\citenamefont {Sasaki}, \citenamefont {Suyama}, \citenamefont {Tanaka},\ and\ \citenamefont {Yokoyama}}]{Sasaki:2016jop}%
  \BibitemOpen
  \bibfield  {author} {\bibinfo {author} {\bibfnamefont {M.}~\bibnamefont {Sasaki}}, \bibinfo {author} {\bibfnamefont {T.}~\bibnamefont {Suyama}}, \bibinfo {author} {\bibfnamefont {T.}~\bibnamefont {Tanaka}}, \ and\ \bibinfo {author} {\bibfnamefont {S.}~\bibnamefont {Yokoyama}},\ }\href {\doibase 10.1103/PhysRevLett.121.059901, 10.1103/PhysRevLett.117.061101} {\bibfield  {journal} {\bibinfo  {journal} {Phys.~Rev.~Lett.}\ }\textbf {\bibinfo {volume} {117}},\ \bibinfo {pages} {061101} (\bibinfo {year} {2016})},\ \bibinfo {note} {[erratum: Phys.~Rev.~Lett.121,no.5,059901(2018)]},\ \Eprint {http://arxiv.org/abs/1603.08338} {arXiv:1603.08338 [astro-ph.CO]} \BibitemShut {NoStop}%
\bibitem [{\citenamefont {Clesse}\ and\ \citenamefont {Garc{\'i}a-Bellido}(2017)}]{Clesse:2016vqa}%
  \BibitemOpen
  \bibfield  {author} {\bibinfo {author} {\bibfnamefont {S.}~\bibnamefont {Clesse}}\ and\ \bibinfo {author} {\bibfnamefont {J.}~\bibnamefont {Garc{\'i}a-Bellido}},\ }\href {\doibase 10.1016/j.dark.2016.10.002} {\bibfield  {journal} {\bibinfo  {journal} {Phys.~Dark Universe}\ }\textbf {\bibinfo {volume} {15}},\ \bibinfo {pages} {142} (\bibinfo {year} {2017})}\BibitemShut {NoStop}%
\bibitem [{\citenamefont {Bird}\ \emph {et~al.}(2016)\citenamefont {Bird}, \citenamefont {Cholis}, \citenamefont {Mu{\~n}oz}, \citenamefont {Ali-Ha{\"i}moud}, \citenamefont {Kamionkowski}, \citenamefont {Kovetz}, \citenamefont {Raccanelli},\ and\ \citenamefont {Riess}}]{Bird:2016dcv}%
  \BibitemOpen
  \bibfield  {author} {\bibinfo {author} {\bibfnamefont {S.}~\bibnamefont {Bird}}, \bibinfo {author} {\bibfnamefont {I.}~\bibnamefont {Cholis}}, \bibinfo {author} {\bibfnamefont {J.~B.}\ \bibnamefont {Mu{\~n}oz}}, \bibinfo {author} {\bibfnamefont {Y.}~\bibnamefont {Ali-Ha{\"i}moud}}, \bibinfo {author} {\bibfnamefont {M.}~\bibnamefont {Kamionkowski}}, \bibinfo {author} {\bibfnamefont {E.~D.}\ \bibnamefont {Kovetz}}, \bibinfo {author} {\bibfnamefont {A.}~\bibnamefont {Raccanelli}}, \ and\ \bibinfo {author} {\bibfnamefont {A.~G.}\ \bibnamefont {Riess}},\ }\href {\doibase 10.1103/PhysRevLett.116.201301} {\bibfield  {journal} {\bibinfo  {journal} {Phys.~Rev.~Lett.}\ }\textbf {\bibinfo {volume} {116}},\ \bibinfo {pages} {201301} (\bibinfo {year} {2016})}\BibitemShut {NoStop}%
\bibitem [{\citenamefont {Allen}\ \emph {et~al.}(2012)\citenamefont {Allen}, \citenamefont {Anderson}, \citenamefont {Brady}, \citenamefont {Brown},\ and\ \citenamefont {Creighton}}]{Allen:2005fk}%
  \BibitemOpen
  \bibfield  {author} {\bibinfo {author} {\bibfnamefont {B.}~\bibnamefont {Allen}}, \bibinfo {author} {\bibfnamefont {W.~G.}\ \bibnamefont {Anderson}}, \bibinfo {author} {\bibfnamefont {P.~R.}\ \bibnamefont {Brady}}, \bibinfo {author} {\bibfnamefont {D.~A.}\ \bibnamefont {Brown}}, \ and\ \bibinfo {author} {\bibfnamefont {J.~D.~E.}\ \bibnamefont {Creighton}},\ }\href {\doibase 10.1103/PhysRevD.85.122006} {\bibfield  {journal} {\bibinfo  {journal} {Phys. Rev. D}\ }\textbf {\bibinfo {volume} {85}},\ \bibinfo {pages} {122006} (\bibinfo {year} {2012})},\ \Eprint {http://arxiv.org/abs/gr-qc/0509116} {arXiv:gr-qc/0509116} \BibitemShut {NoStop}%
\bibitem [{\citenamefont {Dupuis}\ and\ \citenamefont {Woan}(2005)}]{Dupuis:2005xv}%
  \BibitemOpen
  \bibfield  {author} {\bibinfo {author} {\bibfnamefont {R.~J.}\ \bibnamefont {Dupuis}}\ and\ \bibinfo {author} {\bibfnamefont {G.}~\bibnamefont {Woan}},\ }\href {\doibase 10.1103/PhysRevD.72.102002} {\bibfield  {journal} {\bibinfo  {journal} {Phys. Rev. D}\ }\textbf {\bibinfo {volume} {72}},\ \bibinfo {pages} {102002} (\bibinfo {year} {2005})},\ \Eprint {http://arxiv.org/abs/gr-qc/0508096} {arXiv:gr-qc/0508096} \BibitemShut {NoStop}%
\bibitem [{\citenamefont {Maggiore}(2008)}]{maggiore2008gravitational}%
  \BibitemOpen
  \bibfield  {author} {\bibinfo {author} {\bibfnamefont {M.}~\bibnamefont {Maggiore}},\ }\href@noop {} {\emph {\bibinfo {title} {Gravitational Waves: Volume 1: Theory and Experiments}}},\ Vol.~\bibinfo {volume} {1}\ (\bibinfo  {publisher} {Oxford University Press},\ \bibinfo {year} {2008})\BibitemShut {NoStop}%
\bibitem [{\citenamefont {Miller}(2024)}]{Miller:2024rca}%
  \BibitemOpen
  \bibfield  {author} {\bibinfo {author} {\bibfnamefont {A.~L.}\ \bibnamefont {Miller}},\ }\href@noop {} {\  (\bibinfo {year} {2024})},\ \Eprint {http://arxiv.org/abs/2404.11601} {arXiv:2404.11601 [gr-qc]} \BibitemShut {NoStop}%
\bibitem [{\citenamefont {Abbott}\ \emph {et~al.}(2022{\natexlab{a}})\citenamefont {Abbott} \emph {et~al.}}]{LIGOScientific:2021job}%
  \BibitemOpen
  \bibfield  {author} {\bibinfo {author} {\bibfnamefont {R.}~\bibnamefont {Abbott}} \emph {et~al.} (\bibinfo {collaboration} {LIGO Scientific Collaboration, Virgo, KAGRA}),\ }\href {\doibase 10.1103/PhysRevLett.129.061104} {\bibfield  {journal} {\bibinfo  {journal} {Phys. Rev. Lett.}\ }\textbf {\bibinfo {volume} {129}},\ \bibinfo {pages} {061104} (\bibinfo {year} {2022}{\natexlab{a}})},\ \Eprint {http://arxiv.org/abs/2109.12197} {arXiv:2109.12197 [astro-ph.CO]} \BibitemShut {NoStop}%
\bibitem [{\citenamefont {Phukon}\ \emph {et~al.}(2021)\citenamefont {Phukon}, \citenamefont {Baltus}, \citenamefont {Caudill}, \citenamefont {Clesse}, \citenamefont {Depasse}, \citenamefont {Fays}, \citenamefont {Fong}, \citenamefont {Kapadia}, \citenamefont {Magee},\ and\ \citenamefont {Tanasijczuk}}]{Phukon:2021cus}%
  \BibitemOpen
  \bibfield  {author} {\bibinfo {author} {\bibfnamefont {K.~S.}\ \bibnamefont {Phukon}}, \bibinfo {author} {\bibfnamefont {G.}~\bibnamefont {Baltus}}, \bibinfo {author} {\bibfnamefont {S.}~\bibnamefont {Caudill}}, \bibinfo {author} {\bibfnamefont {S.}~\bibnamefont {Clesse}}, \bibinfo {author} {\bibfnamefont {A.}~\bibnamefont {Depasse}}, \bibinfo {author} {\bibfnamefont {M.}~\bibnamefont {Fays}}, \bibinfo {author} {\bibfnamefont {H.}~\bibnamefont {Fong}}, \bibinfo {author} {\bibfnamefont {S.~J.}\ \bibnamefont {Kapadia}}, \bibinfo {author} {\bibfnamefont {R.}~\bibnamefont {Magee}}, \ and\ \bibinfo {author} {\bibfnamefont {A.~J.}\ \bibnamefont {Tanasijczuk}},\ }\href@noop {} {\  (\bibinfo {year} {2021})},\ \Eprint {http://arxiv.org/abs/2105.11449} {arXiv:2105.11449 [astro-ph.CO]} \BibitemShut {NoStop}%
\bibitem [{\citenamefont {Nitz}\ and\ \citenamefont {Wang}(2021)}]{Nitz:2021mzz}%
  \BibitemOpen
  \bibfield  {author} {\bibinfo {author} {\bibfnamefont {A.~H.}\ \bibnamefont {Nitz}}\ and\ \bibinfo {author} {\bibfnamefont {Y.-F.}\ \bibnamefont {Wang}},\ }\href {\doibase 10.3847/1538-4357/ac01d9} {\bibfield  {journal} {\bibinfo  {journal} {The Astrophysical Journal}\ }\textbf {\bibinfo {volume} {915}},\ \bibinfo {pages} {54} (\bibinfo {year} {2021})},\ \Eprint {http://arxiv.org/abs/arXiv:2102.00868} {arXiv:2102.00868} \BibitemShut {NoStop}%
\bibitem [{\citenamefont {Nitz}\ and\ \citenamefont {Wang}(2022)}]{Nitz:2022ltl}%
  \BibitemOpen
  \bibfield  {author} {\bibinfo {author} {\bibfnamefont {A.~H.}\ \bibnamefont {Nitz}}\ and\ \bibinfo {author} {\bibfnamefont {Y.-F.}\ \bibnamefont {Wang}},\ }\href {\doibase 10.1103/PhysRevD.106.023024} {\bibfield  {journal} {\bibinfo  {journal} {Phys. Rev. D}\ }\textbf {\bibinfo {volume} {106}},\ \bibinfo {pages} {023024} (\bibinfo {year} {2022})},\ \Eprint {http://arxiv.org/abs/2202.11024} {arXiv:2202.11024 [astro-ph.HE]} \BibitemShut {NoStop}%
\bibitem [{\citenamefont {Bandopadhyay}\ \emph {et~al.}(2022)\citenamefont {Bandopadhyay}, \citenamefont {Reed}, \citenamefont {Padamata}, \citenamefont {Leon}, \citenamefont {Horowitz}, \citenamefont {Brown}, \citenamefont {Radice}, \citenamefont {Fattoyev},\ and\ \citenamefont {Piekarewicz}}]{Bandopadhyay:2022tbi}%
  \BibitemOpen
  \bibfield  {author} {\bibinfo {author} {\bibfnamefont {A.}~\bibnamefont {Bandopadhyay}}, \bibinfo {author} {\bibfnamefont {B.}~\bibnamefont {Reed}}, \bibinfo {author} {\bibfnamefont {S.}~\bibnamefont {Padamata}}, \bibinfo {author} {\bibfnamefont {E.}~\bibnamefont {Leon}}, \bibinfo {author} {\bibfnamefont {C.~J.}\ \bibnamefont {Horowitz}}, \bibinfo {author} {\bibfnamefont {D.~A.}\ \bibnamefont {Brown}}, \bibinfo {author} {\bibfnamefont {D.}~\bibnamefont {Radice}}, \bibinfo {author} {\bibfnamefont {F.~J.}\ \bibnamefont {Fattoyev}}, \ and\ \bibinfo {author} {\bibfnamefont {J.}~\bibnamefont {Piekarewicz}},\ }\href@noop {} {\  (\bibinfo {year} {2022})},\ \Eprint {http://arxiv.org/abs/2212.03855} {arXiv:2212.03855 [astro-ph.HE]} \BibitemShut {NoStop}%
\bibitem [{\citenamefont {Abbott}\ \emph {et~al.}(2023)\citenamefont {Abbott} \emph {et~al.}}]{LIGOScientific:2022hai}%
  \BibitemOpen
  \bibfield  {author} {\bibinfo {author} {\bibfnamefont {R.}~\bibnamefont {Abbott}} \emph {et~al.} (\bibinfo {collaboration} {LIGO Scientific, VIRGO, KAGRA}),\ }\href {\doibase 10.1093/mnras/stad588} {\bibfield  {journal} {\bibinfo  {journal} {Mon. Not. Roy. Astron. Soc.}\ }\textbf {\bibinfo {volume} {524}},\ \bibinfo {pages} {5984} (\bibinfo {year} {2023})},\ \bibinfo {note} {[Erratum: Mon.Not.Roy.Astron.Soc. 526, 6234 (2023)]},\ \Eprint {http://arxiv.org/abs/2212.01477} {arXiv:2212.01477 [astro-ph.HE]} \BibitemShut {NoStop}%
\bibitem [{\citenamefont {Miller}\ \emph {et~al.}(2024{\natexlab{b}})\citenamefont {Miller}, \citenamefont {Singh},\ and\ \citenamefont {Palomba}}]{Miller:2023rnn}%
  \BibitemOpen
  \bibfield  {author} {\bibinfo {author} {\bibfnamefont {A.~L.}\ \bibnamefont {Miller}}, \bibinfo {author} {\bibfnamefont {N.}~\bibnamefont {Singh}}, \ and\ \bibinfo {author} {\bibfnamefont {C.}~\bibnamefont {Palomba}},\ }\href {\doibase 10.1103/PhysRevD.109.043021} {\bibfield  {journal} {\bibinfo  {journal} {Phys. Rev. D}\ }\textbf {\bibinfo {volume} {109}},\ \bibinfo {pages} {043021} (\bibinfo {year} {2024}{\natexlab{b}})},\ \Eprint {http://arxiv.org/abs/2309.15808} {arXiv:2309.15808 [astro-ph.IM]} \BibitemShut {NoStop}%
\bibitem [{\citenamefont {Miller}\ \emph {et~al.}(2022)\citenamefont {Miller}, \citenamefont {Aggarwal}, \citenamefont {Clesse},\ and\ \citenamefont {De~Lillo}}]{Miller:2021knj}%
  \BibitemOpen
  \bibfield  {author} {\bibinfo {author} {\bibfnamefont {A.~L.}\ \bibnamefont {Miller}}, \bibinfo {author} {\bibfnamefont {N.}~\bibnamefont {Aggarwal}}, \bibinfo {author} {\bibfnamefont {S.}~\bibnamefont {Clesse}}, \ and\ \bibinfo {author} {\bibfnamefont {F.}~\bibnamefont {De~Lillo}},\ }\href {\doibase 10.1103/PhysRevD.105.062008} {\bibfield  {journal} {\bibinfo  {journal} {Phys. Rev. D}\ }\textbf {\bibinfo {volume} {105}},\ \bibinfo {pages} {062008} (\bibinfo {year} {2022})},\ \Eprint {http://arxiv.org/abs/2110.06188} {arXiv:2110.06188 [gr-qc]} \BibitemShut {NoStop}%
\bibitem [{\citenamefont {Astone}\ \emph {et~al.}(2014)\citenamefont {Astone}, \citenamefont {Colla}, \citenamefont {D'Antonio}, \citenamefont {Frasca},\ and\ \citenamefont {Palomba}}]{Astone:2014esa}%
  \BibitemOpen
  \bibfield  {author} {\bibinfo {author} {\bibfnamefont {P.}~\bibnamefont {Astone}}, \bibinfo {author} {\bibfnamefont {A.}~\bibnamefont {Colla}}, \bibinfo {author} {\bibfnamefont {S.}~\bibnamefont {D'Antonio}}, \bibinfo {author} {\bibfnamefont {S.}~\bibnamefont {Frasca}}, \ and\ \bibinfo {author} {\bibfnamefont {C.}~\bibnamefont {Palomba}},\ }\href@noop {} {\bibfield  {journal} {\bibinfo  {journal} {Physical Review D}\ }\textbf {\bibinfo {volume} {90}},\ \bibinfo {pages} {042002} (\bibinfo {year} {2014})}\BibitemShut {NoStop}%
\bibitem [{\citenamefont {Miller}\ \emph {et~al.}(2018)\citenamefont {Miller} \emph {et~al.}}]{Miller:2018rbg}%
  \BibitemOpen
  \bibfield  {author} {\bibinfo {author} {\bibfnamefont {A.}~\bibnamefont {Miller}} \emph {et~al.},\ }\href {\doibase 10.1103/PhysRevD.98.102004} {\bibfield  {journal} {\bibinfo  {journal} {Phys. Rev. D}\ }\textbf {\bibinfo {volume} {98}},\ \bibinfo {pages} {102004} (\bibinfo {year} {2018})},\ \Eprint {http://arxiv.org/abs/1810.09784} {arXiv:1810.09784 [astro-ph.IM]} \BibitemShut {NoStop}%
\bibitem [{\citenamefont {Miller}\ \emph {et~al.}(2021)\citenamefont {Miller}, \citenamefont {Clesse}, \citenamefont {De~Lillo}, \citenamefont {Bruno}, \citenamefont {Depasse},\ and\ \citenamefont {Tanasijczuk}}]{Miller:2020kmv}%
  \BibitemOpen
  \bibfield  {author} {\bibinfo {author} {\bibfnamefont {A.~L.}\ \bibnamefont {Miller}}, \bibinfo {author} {\bibfnamefont {S.}~\bibnamefont {Clesse}}, \bibinfo {author} {\bibfnamefont {F.}~\bibnamefont {De~Lillo}}, \bibinfo {author} {\bibfnamefont {G.}~\bibnamefont {Bruno}}, \bibinfo {author} {\bibfnamefont {A.}~\bibnamefont {Depasse}}, \ and\ \bibinfo {author} {\bibfnamefont {A.}~\bibnamefont {Tanasijczuk}},\ }\href {\doibase 10.1016/j.dark.2021.100836} {\bibfield  {journal} {\bibinfo  {journal} {Phys. Dark Univ.}\ }\textbf {\bibinfo {volume} {32}},\ \bibinfo {pages} {100836} (\bibinfo {year} {2021})},\ \Eprint {http://arxiv.org/abs/2012.12983} {arXiv:2012.12983 [astro-ph.HE]} \BibitemShut {NoStop}%
\bibitem [{\citenamefont {Andrés-Carcasona}\ \emph {et~al.}(2023)\citenamefont {Andrés-Carcasona}, \citenamefont {Piccinni}, \citenamefont {Martínez},\ and\ \citenamefont {Mir}}]{Andrés-Carcasona:2023df}%
  \BibitemOpen
  \bibfield  {author} {\bibinfo {author} {\bibfnamefont {M.}~\bibnamefont {Andrés-Carcasona}}, \bibinfo {author} {\bibfnamefont {O.~J.}\ \bibnamefont {Piccinni}}, \bibinfo {author} {\bibfnamefont {M.}~\bibnamefont {Martínez}}, \ and\ \bibinfo {author} {\bibfnamefont {L.-M.}\ \bibnamefont {Mir}},\ }\href {\doibase 10.22323/1.449.0067} {\bibfield  {journal} {\bibinfo  {journal} {PoS}\ }\textbf {\bibinfo {volume} {EPS-HEP2023}},\ \bibinfo {pages} {067} (\bibinfo {year} {2023})}\BibitemShut {NoStop}%
\bibitem [{\citenamefont {Alestas}\ \emph {et~al.}(2024)\citenamefont {Alestas}, \citenamefont {Morras}, \citenamefont {Yamamoto}, \citenamefont {Garcia-Bellido}, \citenamefont {Kuroyanagi},\ and\ \citenamefont {Nesseris}}]{Alestas:2024ubs}%
  \BibitemOpen
  \bibfield  {author} {\bibinfo {author} {\bibfnamefont {G.}~\bibnamefont {Alestas}}, \bibinfo {author} {\bibfnamefont {G.}~\bibnamefont {Morras}}, \bibinfo {author} {\bibfnamefont {T.~S.}\ \bibnamefont {Yamamoto}}, \bibinfo {author} {\bibfnamefont {J.}~\bibnamefont {Garcia-Bellido}}, \bibinfo {author} {\bibfnamefont {S.}~\bibnamefont {Kuroyanagi}}, \ and\ \bibinfo {author} {\bibfnamefont {S.}~\bibnamefont {Nesseris}},\ }\href {\doibase 10.1103/PhysRevD.109.123516} {\bibfield  {journal} {\bibinfo  {journal} {Phys. Rev. D}\ }\textbf {\bibinfo {volume} {109}},\ \bibinfo {pages} {123516} (\bibinfo {year} {2024})},\ \Eprint {http://arxiv.org/abs/2401.02314} {arXiv:2401.02314 [astro-ph.CO]} \BibitemShut {NoStop}%
\bibitem [{\citenamefont {Sun}\ and\ \citenamefont {Melatos}(2019)}]{Sun:2018hmm}%
  \BibitemOpen
  \bibfield  {author} {\bibinfo {author} {\bibfnamefont {L.}~\bibnamefont {Sun}}\ and\ \bibinfo {author} {\bibfnamefont {A.}~\bibnamefont {Melatos}},\ }\href@noop {} {\bibfield  {journal} {\bibinfo  {journal} {Physical Review D}\ }\textbf {\bibinfo {volume} {99}},\ \bibinfo {pages} {123003} (\bibinfo {year} {2019})}\BibitemShut {NoStop}%
\bibitem [{\citenamefont {Oliver}\ \emph {et~al.}(2019)\citenamefont {Oliver}, \citenamefont {Keitel},\ and\ \citenamefont {Sintes}}]{Oliver:2018dpt}%
  \BibitemOpen
  \bibfield  {author} {\bibinfo {author} {\bibfnamefont {M.}~\bibnamefont {Oliver}}, \bibinfo {author} {\bibfnamefont {D.}~\bibnamefont {Keitel}}, \ and\ \bibinfo {author} {\bibfnamefont {A.~M.}\ \bibnamefont {Sintes}},\ }\href@noop {} {\bibfield  {journal} {\bibinfo  {journal} {Physical Review D}\ }\textbf {\bibinfo {volume} {99}},\ \bibinfo {pages} {104067} (\bibinfo {year} {2019})}\BibitemShut {NoStop}%
\bibitem [{\citenamefont {Miller}\ \emph {et~al.}(2019)\citenamefont {Miller} \emph {et~al.}}]{Miller:2019jtp}%
  \BibitemOpen
  \bibfield  {author} {\bibinfo {author} {\bibfnamefont {A.~L.}\ \bibnamefont {Miller}} \emph {et~al.},\ }\href {\doibase 10.1103/PhysRevD.100.062005} {\bibfield  {journal} {\bibinfo  {journal} {Phys. Rev. D}\ }\textbf {\bibinfo {volume} {100}},\ \bibinfo {pages} {062005} (\bibinfo {year} {2019})},\ \Eprint {http://arxiv.org/abs/1909.02262} {arXiv:1909.02262 [astro-ph.IM]} \BibitemShut {NoStop}%
\bibitem [{\citenamefont {Abbott}\ \emph {et~al.}(2019{\natexlab{a}})\citenamefont {Abbott} \emph {et~al.}}]{LIGOScientific:2019kan}%
  \BibitemOpen
  \bibfield  {author} {\bibinfo {author} {\bibfnamefont {B.~P.}\ \bibnamefont {Abbott}} \emph {et~al.} (\bibinfo {collaboration} {LIGO Scientific Collaboration, Virgo}),\ }\href {\doibase 10.1103/PhysRevLett.123.161102} {\bibfield  {journal} {\bibinfo  {journal} {Phys. Rev. Lett.}\ }\textbf {\bibinfo {volume} {123}},\ \bibinfo {pages} {161102} (\bibinfo {year} {2019}{\natexlab{a}})},\ \Eprint {http://arxiv.org/abs/1904.08976} {arXiv:1904.08976 [astro-ph.CO]} \BibitemShut {NoStop}%
\bibitem [{\citenamefont {Cannon}\ \emph {et~al.}(2012)\citenamefont {Cannon} \emph {et~al.}}]{Cannon:2011vi}%
  \BibitemOpen
  \bibfield  {author} {\bibinfo {author} {\bibfnamefont {K.}~\bibnamefont {Cannon}} \emph {et~al.},\ }\href {\doibase 10.1088/0004-637X/748/2/136} {\bibfield  {journal} {\bibinfo  {journal} {Astrophys. J.}\ }\textbf {\bibinfo {volume} {748}},\ \bibinfo {pages} {136} (\bibinfo {year} {2012})},\ \Eprint {http://arxiv.org/abs/1107.2665} {arXiv:1107.2665 [astro-ph.IM]} \BibitemShut {NoStop}%
\bibitem [{\citenamefont {Zhao}\ and\ \citenamefont {Wen}(2018)}]{Zhao:2017cbb}%
  \BibitemOpen
  \bibfield  {author} {\bibinfo {author} {\bibfnamefont {W.}~\bibnamefont {Zhao}}\ and\ \bibinfo {author} {\bibfnamefont {L.}~\bibnamefont {Wen}},\ }\href {\doibase 10.1103/PhysRevD.97.064031} {\bibfield  {journal} {\bibinfo  {journal} {Phys. Rev. D}\ }\textbf {\bibinfo {volume} {97}},\ \bibinfo {pages} {064031} (\bibinfo {year} {2018})},\ \Eprint {http://arxiv.org/abs/1710.05325} {arXiv:1710.05325 [astro-ph.CO]} \BibitemShut {NoStop}%
\bibitem [{\citenamefont {Chan}\ \emph {et~al.}(2018)\citenamefont {Chan}, \citenamefont {Messenger}, \citenamefont {Heng},\ and\ \citenamefont {Hendry}}]{Chan:2018csa}%
  \BibitemOpen
  \bibfield  {author} {\bibinfo {author} {\bibfnamefont {M.~L.}\ \bibnamefont {Chan}}, \bibinfo {author} {\bibfnamefont {C.}~\bibnamefont {Messenger}}, \bibinfo {author} {\bibfnamefont {I.~S.}\ \bibnamefont {Heng}}, \ and\ \bibinfo {author} {\bibfnamefont {M.}~\bibnamefont {Hendry}},\ }\href {\doibase 10.1103/PhysRevD.97.123014} {\bibfield  {journal} {\bibinfo  {journal} {Phys. Rev. D}\ }\textbf {\bibinfo {volume} {97}},\ \bibinfo {pages} {123014} (\bibinfo {year} {2018})},\ \Eprint {http://arxiv.org/abs/1803.09680} {arXiv:1803.09680 [astro-ph.HE]} \BibitemShut {NoStop}%
\bibitem [{\citenamefont {Sachdev}\ \emph {et~al.}(2020)\citenamefont {Sachdev} \emph {et~al.}}]{Sachdev:2020lfd}%
  \BibitemOpen
  \bibfield  {author} {\bibinfo {author} {\bibfnamefont {S.}~\bibnamefont {Sachdev}} \emph {et~al.},\ }\href {\doibase 10.3847/2041-8213/abc753} {\bibfield  {journal} {\bibinfo  {journal} {Astrophys. J. Lett.}\ }\textbf {\bibinfo {volume} {905}},\ \bibinfo {pages} {L25} (\bibinfo {year} {2020})},\ \Eprint {http://arxiv.org/abs/2008.04288} {arXiv:2008.04288 [astro-ph.HE]} \BibitemShut {NoStop}%
\bibitem [{\citenamefont {Banerjee}\ \emph {et~al.}(2023)\citenamefont {Banerjee} \emph {et~al.}}]{Banerjee:2022gkv}%
  \BibitemOpen
  \bibfield  {author} {\bibinfo {author} {\bibfnamefont {B.}~\bibnamefont {Banerjee}} \emph {et~al.},\ }\href {\doibase 10.1051/0004-6361/202345850} {\bibfield  {journal} {\bibinfo  {journal} {Astron. Astrophys.}\ }\textbf {\bibinfo {volume} {678}},\ \bibinfo {pages} {A126} (\bibinfo {year} {2023})},\ \Eprint {http://arxiv.org/abs/2212.14007} {arXiv:2212.14007 [astro-ph.HE]} \BibitemShut {NoStop}%
\bibitem [{\citenamefont {Nitz}\ and\ \citenamefont {Dal~Canton}(2021)}]{Nitz:2021pbr}%
  \BibitemOpen
  \bibfield  {author} {\bibinfo {author} {\bibfnamefont {A.~H.}\ \bibnamefont {Nitz}}\ and\ \bibinfo {author} {\bibfnamefont {T.}~\bibnamefont {Dal~Canton}},\ }\href {\doibase 10.3847/2041-8213/ac1a75} {\bibfield  {journal} {\bibinfo  {journal} {Astrophys. J. Lett.}\ }\textbf {\bibinfo {volume} {917}},\ \bibinfo {pages} {L27} (\bibinfo {year} {2021})},\ \Eprint {http://arxiv.org/abs/2106.15259} {arXiv:2106.15259 [astro-ph.HE]} \BibitemShut {NoStop}%
\bibitem [{\citenamefont {Baral}\ \emph {et~al.}(2023)\citenamefont {Baral}, \citenamefont {Morisaki}, \citenamefont {Maga\~na Hernandez},\ and\ \citenamefont {Creighton}}]{Baral:2023xst}%
  \BibitemOpen
  \bibfield  {author} {\bibinfo {author} {\bibfnamefont {P.}~\bibnamefont {Baral}}, \bibinfo {author} {\bibfnamefont {S.}~\bibnamefont {Morisaki}}, \bibinfo {author} {\bibfnamefont {I.}~\bibnamefont {Maga\~na Hernandez}}, \ and\ \bibinfo {author} {\bibfnamefont {J.}~\bibnamefont {Creighton}},\ }\href {\doibase 10.1103/PhysRevD.108.043010} {\bibfield  {journal} {\bibinfo  {journal} {Phys. Rev. D}\ }\textbf {\bibinfo {volume} {108}},\ \bibinfo {pages} {043010} (\bibinfo {year} {2023})},\ \Eprint {http://arxiv.org/abs/2304.09889} {arXiv:2304.09889 [astro-ph.HE]} \BibitemShut {NoStop}%
\bibitem [{\citenamefont {Astone}\ \emph {et~al.}(2005)\citenamefont {Astone}, \citenamefont {Frasca},\ and\ \citenamefont {Palomba}}]{Astone:2005fj}%
  \BibitemOpen
  \bibfield  {author} {\bibinfo {author} {\bibfnamefont {P.}~\bibnamefont {Astone}}, \bibinfo {author} {\bibfnamefont {S.}~\bibnamefont {Frasca}}, \ and\ \bibinfo {author} {\bibfnamefont {C.}~\bibnamefont {Palomba}},\ }\href {\doibase 10.1088/0264-9381/22/18/S34} {\bibfield  {journal} {\bibinfo  {journal} {Class. Quant. Grav.}\ }\textbf {\bibinfo {volume} {22}},\ \bibinfo {pages} {S1197} (\bibinfo {year} {2005})}\BibitemShut {NoStop}%
\bibitem [{\citenamefont {Acernese}\ \emph {et~al.}(2009)\citenamefont {Acernese} \emph {et~al.}}]{Acernese:2009zz}%
  \BibitemOpen
  \bibfield  {author} {\bibinfo {author} {\bibfnamefont {F.}~\bibnamefont {Acernese}} \emph {et~al.},\ }\href {\doibase 10.1088/0264-9381/26/20/204002} {\bibfield  {journal} {\bibinfo  {journal} {Class. Quant. Grav.}\ }\textbf {\bibinfo {volume} {26}},\ \bibinfo {pages} {204002} (\bibinfo {year} {2009})}\BibitemShut {NoStop}%
\bibitem [{\citenamefont {Pierini}\ and\ \citenamefont {Felicetti}(2022)}]{lorenzoposter}%
  \BibitemOpen
  \bibfield  {author} {\bibinfo {author} {\bibfnamefont {L.}~\bibnamefont {Pierini}}\ and\ \bibinfo {author} {\bibfnamefont {R.}~\bibnamefont {Felicetti}},\ }\href {https://k-poster.kuoni-congress.info/eas2022/poster/7e6eec03-5b5d-4dd7-b24e-13b3c6bb0067} {\enquote {\bibinfo {title} {{Continuous gravitational waves from scalar boson clouds around kerr black holes: obesrvational issues and search strategies}},}\ } (\bibinfo {year} {2022})\BibitemShut {NoStop}%
\bibitem [{\citenamefont {Miller}\ \emph {et~al.}(2024{\natexlab{c}})\citenamefont {Miller} \emph {et~al.}}]{miller_2024_10724845}%
  \BibitemOpen
  \bibfield  {author} {\bibinfo {author} {\bibfnamefont {A.~L.}\ \bibnamefont {Miller}} \emph {et~al.},\ }\href {\doibase https://doi.org/10.5281/zenodo.10724845} {\bibfield  {journal} {\bibinfo  {journal} {Data release for Gravitational wave constraints on planetary-mass primordial black holes using LIGO O3a data, Zenodo}\ } (\bibinfo {year} {2024}{\natexlab{c}}),\ https://doi.org/10.5281/zenodo.10724845}\BibitemShut {NoStop}%
\bibitem [{\citenamefont {Sun}\ \emph {et~al.}(2020)\citenamefont {Sun} \emph {et~al.}}]{Sun:2020wke}%
  \BibitemOpen
  \bibfield  {author} {\bibinfo {author} {\bibfnamefont {L.}~\bibnamefont {Sun}} \emph {et~al.},\ }\href {\doibase 10.1088/1361-6382/abb14e} {\bibfield  {journal} {\bibinfo  {journal} {Class. Quant. Grav.}\ }\textbf {\bibinfo {volume} {37}},\ \bibinfo {pages} {225008} (\bibinfo {year} {2020})},\ \Eprint {http://arxiv.org/abs/2005.02531} {arXiv:2005.02531 [astro-ph.IM]} \BibitemShut {NoStop}%
\bibitem [{\citenamefont {Abbott}\ \emph {et~al.}(2019{\natexlab{b}})\citenamefont {Abbott} \emph {et~al.}}]{LIGOScientific:2018urg}%
  \BibitemOpen
  \bibfield  {author} {\bibinfo {author} {\bibfnamefont {B.~P.}\ \bibnamefont {Abbott}} \emph {et~al.} (\bibinfo {collaboration} {LIGO Scientific Collaboration, Virgo}),\ }\href {\doibase 10.3847/1538-4357/ab0f3d} {\bibfield  {journal} {\bibinfo  {journal} {Astrophys. J.}\ }\textbf {\bibinfo {volume} {875}},\ \bibinfo {pages} {160} (\bibinfo {year} {2019}{\natexlab{b}})},\ \Eprint {http://arxiv.org/abs/1810.02581} {arXiv:1810.02581 [gr-qc]} \BibitemShut {NoStop}%
\bibitem [{\citenamefont {Piccinni}\ \emph {et~al.}(2019)\citenamefont {Piccinni}, \citenamefont {Astone}, \citenamefont {D'Antonio}, \citenamefont {Frasca}, \citenamefont {Intini}, \citenamefont {Leaci}, \citenamefont {Mastrogiovanni}, \citenamefont {Miller}, \citenamefont {Palomba},\ and\ \citenamefont {Singhal}}]{Piccinni:2018akm}%
  \BibitemOpen
  \bibfield  {author} {\bibinfo {author} {\bibfnamefont {O.~J.}\ \bibnamefont {Piccinni}}, \bibinfo {author} {\bibfnamefont {P.}~\bibnamefont {Astone}}, \bibinfo {author} {\bibfnamefont {S.}~\bibnamefont {D'Antonio}}, \bibinfo {author} {\bibfnamefont {S.}~\bibnamefont {Frasca}}, \bibinfo {author} {\bibfnamefont {G.}~\bibnamefont {Intini}}, \bibinfo {author} {\bibfnamefont {P.}~\bibnamefont {Leaci}}, \bibinfo {author} {\bibfnamefont {S.}~\bibnamefont {Mastrogiovanni}}, \bibinfo {author} {\bibfnamefont {A.}~\bibnamefont {Miller}}, \bibinfo {author} {\bibfnamefont {C.}~\bibnamefont {Palomba}}, \ and\ \bibinfo {author} {\bibfnamefont {A.}~\bibnamefont {Singhal}},\ }\href {\doibase 10.1088/1361-6382/aaefb5} {\bibfield  {journal} {\bibinfo  {journal} {Class. Quant. Grav.}\ }\textbf {\bibinfo {volume} {36}},\ \bibinfo {pages} {015008} (\bibinfo {year} {2019})},\ \Eprint {http://arxiv.org/abs/1811.04730} {arXiv:1811.04730 [gr-qc]} \BibitemShut {NoStop}%
\bibitem [{\citenamefont {Abbott}\ \emph {et~al.}(2020)\citenamefont {Abbott} \emph {et~al.}}]{LIGOScientific:2020gml}%
  \BibitemOpen
  \bibfield  {author} {\bibinfo {author} {\bibfnamefont {R.}~\bibnamefont {Abbott}} \emph {et~al.} (\bibinfo {collaboration} {LIGO Scientific Collaboration, Virgo}),\ }\href {\doibase 10.3847/2041-8213/abb655} {\bibfield  {journal} {\bibinfo  {journal} {Astrophys. J. Lett.}\ }\textbf {\bibinfo {volume} {902}},\ \bibinfo {pages} {L21} (\bibinfo {year} {2020})},\ \Eprint {http://arxiv.org/abs/2007.14251} {arXiv:2007.14251 [astro-ph.HE]} \BibitemShut {NoStop}%
\bibitem [{\citenamefont {Abbott}\ \emph {et~al.}(2022{\natexlab{b}})\citenamefont {Abbott} \emph {et~al.}}]{LIGOScientific:2021hvc}%
  \BibitemOpen
  \bibfield  {author} {\bibinfo {author} {\bibfnamefont {R.}~\bibnamefont {Abbott}} \emph {et~al.} (\bibinfo {collaboration} {LIGO Scientific Collaboration, Virgo, KAGRA}),\ }\href {\doibase 10.3847/1538-4357/ac6acf} {\bibfield  {journal} {\bibinfo  {journal} {Astrophys. J.}\ }\textbf {\bibinfo {volume} {935}},\ \bibinfo {pages} {1} (\bibinfo {year} {2022}{\natexlab{b}})},\ \Eprint {http://arxiv.org/abs/2111.13106} {arXiv:2111.13106 [astro-ph.HE]} \BibitemShut {NoStop}%
\bibitem [{\citenamefont {Abbott}\ \emph {et~al.}(2021)\citenamefont {Abbott} \emph {et~al.}}]{LIGOScientific:2020ibl}%
  \BibitemOpen
  \bibfield  {author} {\bibinfo {author} {\bibfnamefont {R.}~\bibnamefont {Abbott}} \emph {et~al.} (\bibinfo {collaboration} {LIGO Scientific Collaboration, Virgo}),\ }\href {\doibase 10.1103/PhysRevX.11.021053} {\bibfield  {journal} {\bibinfo  {journal} {Phys. Rev. X}\ }\textbf {\bibinfo {volume} {11}},\ \bibinfo {pages} {021053} (\bibinfo {year} {2021})},\ \Eprint {http://arxiv.org/abs/2010.14527} {arXiv:2010.14527 [gr-qc]} \BibitemShut {NoStop}%
\bibitem [{\citenamefont {Dreissigacker}\ \emph {et~al.}(2019)\citenamefont {Dreissigacker}, \citenamefont {Sharma}, \citenamefont {Messenger}, \citenamefont {Zhao},\ and\ \citenamefont {Prix}}]{Dreissigacker:2019edy}%
  \BibitemOpen
  \bibfield  {author} {\bibinfo {author} {\bibfnamefont {C.}~\bibnamefont {Dreissigacker}}, \bibinfo {author} {\bibfnamefont {R.}~\bibnamefont {Sharma}}, \bibinfo {author} {\bibfnamefont {C.}~\bibnamefont {Messenger}}, \bibinfo {author} {\bibfnamefont {R.}~\bibnamefont {Zhao}}, \ and\ \bibinfo {author} {\bibfnamefont {R.}~\bibnamefont {Prix}},\ }\href {\doibase 10.1103/PhysRevD.100.044009} {\bibfield  {journal} {\bibinfo  {journal} {Phys. Rev. D}\ }\textbf {\bibinfo {volume} {100}},\ \bibinfo {pages} {044009} (\bibinfo {year} {2019})},\ \Eprint {http://arxiv.org/abs/1904.13291} {arXiv:1904.13291 [gr-qc]} \BibitemShut {NoStop}%
\bibitem [{\citenamefont {Bayley}\ \emph {et~al.}(2022)\citenamefont {Bayley}, \citenamefont {Messenger},\ and\ \citenamefont {Woan}}]{Bayley:2022hkz}%
  \BibitemOpen
  \bibfield  {author} {\bibinfo {author} {\bibfnamefont {J.}~\bibnamefont {Bayley}}, \bibinfo {author} {\bibfnamefont {C.}~\bibnamefont {Messenger}}, \ and\ \bibinfo {author} {\bibfnamefont {G.}~\bibnamefont {Woan}},\ }\href {\doibase 10.1103/PhysRevD.106.083022} {\bibfield  {journal} {\bibinfo  {journal} {Phys. Rev. D}\ }\textbf {\bibinfo {volume} {106}},\ \bibinfo {pages} {083022} (\bibinfo {year} {2022})},\ \Eprint {http://arxiv.org/abs/2209.02031} {arXiv:2209.02031 [astro-ph.IM]} \BibitemShut {NoStop}%
\bibitem [{\citenamefont {Yamamoto}\ \emph {et~al.}(2022)\citenamefont {Yamamoto}, \citenamefont {Miller}, \citenamefont {Sieniawska},\ and\ \citenamefont {Tanaka}}]{Yamamoto:2022adl}%
  \BibitemOpen
  \bibfield  {author} {\bibinfo {author} {\bibfnamefont {T.~S.}\ \bibnamefont {Yamamoto}}, \bibinfo {author} {\bibfnamefont {A.~L.}\ \bibnamefont {Miller}}, \bibinfo {author} {\bibfnamefont {M.}~\bibnamefont {Sieniawska}}, \ and\ \bibinfo {author} {\bibfnamefont {T.}~\bibnamefont {Tanaka}},\ }\href {\doibase 10.1103/PhysRevD.106.024025} {\bibfield  {journal} {\bibinfo  {journal} {Phys. Rev. D}\ }\textbf {\bibinfo {volume} {106}},\ \bibinfo {pages} {024025} (\bibinfo {year} {2022})},\ \Eprint {http://arxiv.org/abs/2206.00882} {arXiv:2206.00882 [gr-qc]} \BibitemShut {NoStop}%
\bibitem [{\citenamefont {Abbott}\ \emph {et~al.}(2022{\natexlab{c}})\citenamefont {Abbott} \emph {et~al.}}]{KAGRA:2022dwb}%
  \BibitemOpen
  \bibfield  {author} {\bibinfo {author} {\bibfnamefont {R.}~\bibnamefont {Abbott}} \emph {et~al.} (\bibinfo {collaboration} {LIGO Scientific Collaboration, Virgo, KAGRA}),\ }\href {\doibase 10.1103/PhysRevD.106.102008} {\bibfield  {journal} {\bibinfo  {journal} {Phys. Rev. D}\ }\textbf {\bibinfo {volume} {106}},\ \bibinfo {pages} {102008} (\bibinfo {year} {2022}{\natexlab{c}})},\ \Eprint {http://arxiv.org/abs/2201.00697} {arXiv:2201.00697 [gr-qc]} \BibitemShut {NoStop}%
\bibitem [{\citenamefont {Steltner}\ \emph {et~al.}(2023)\citenamefont {Steltner}, \citenamefont {Papa}, \citenamefont {Eggenstein}, \citenamefont {Prix}, \citenamefont {Bensch}, \citenamefont {Allen},\ and\ \citenamefont {Machenschalk}}]{Steltner:2023cfk}%
  \BibitemOpen
  \bibfield  {author} {\bibinfo {author} {\bibfnamefont {B.}~\bibnamefont {Steltner}}, \bibinfo {author} {\bibfnamefont {M.~A.}\ \bibnamefont {Papa}}, \bibinfo {author} {\bibfnamefont {H.~B.}\ \bibnamefont {Eggenstein}}, \bibinfo {author} {\bibfnamefont {R.}~\bibnamefont {Prix}}, \bibinfo {author} {\bibfnamefont {M.}~\bibnamefont {Bensch}}, \bibinfo {author} {\bibfnamefont {B.}~\bibnamefont {Allen}}, \ and\ \bibinfo {author} {\bibfnamefont {B.}~\bibnamefont {Machenschalk}},\ }\href {\doibase 10.3847/1538-4357/acdad4} {\bibfield  {journal} {\bibinfo  {journal} {Astrophys. J.}\ }\textbf {\bibinfo {volume} {952}},\ \bibinfo {pages} {55} (\bibinfo {year} {2023})},\ \Eprint {http://arxiv.org/abs/2303.04109} {arXiv:2303.04109 [gr-qc]} \BibitemShut {NoStop}%
\bibitem [{\citenamefont {Abbott}\ \emph {et~al.}(2019{\natexlab{c}})\citenamefont {Abbott} \emph {et~al.}}]{LIGOScientific:2019yhl}%
  \BibitemOpen
  \bibfield  {author} {\bibinfo {author} {\bibfnamefont {B.~P.}\ \bibnamefont {Abbott}} \emph {et~al.} (\bibinfo {collaboration} {LIGO Scientific Collaboration, Virgo}),\ }\href {\doibase 10.1103/PhysRevD.100.024004} {\bibfield  {journal} {\bibinfo  {journal} {Phys. Rev. D}\ }\textbf {\bibinfo {volume} {100}},\ \bibinfo {pages} {024004} (\bibinfo {year} {2019}{\natexlab{c}})},\ \Eprint {http://arxiv.org/abs/1903.01901} {arXiv:1903.01901 [astro-ph.HE]} \BibitemShut {NoStop}%
\bibitem [{\citenamefont {Guo}\ and\ \citenamefont {Miller}(2022)}]{Guo:2022sdd}%
  \BibitemOpen
  \bibfield  {author} {\bibinfo {author} {\bibfnamefont {H.-K.}\ \bibnamefont {Guo}}\ and\ \bibinfo {author} {\bibfnamefont {A.}~\bibnamefont {Miller}},\ }\href@noop {} {\  (\bibinfo {year} {2022})},\ \Eprint {http://arxiv.org/abs/2205.10359} {arXiv:2205.10359 [astro-ph.IM]} \BibitemShut {NoStop}%
\bibitem [{\citenamefont {Carr}\ \emph {et~al.}(2021)\citenamefont {Carr}, \citenamefont {Clesse}, \citenamefont {Garc\'\i{}a-Bellido},\ and\ \citenamefont {K\"uhnel}}]{Carr:2019kxo}%
  \BibitemOpen
  \bibfield  {author} {\bibinfo {author} {\bibfnamefont {B.}~\bibnamefont {Carr}}, \bibinfo {author} {\bibfnamefont {S.}~\bibnamefont {Clesse}}, \bibinfo {author} {\bibfnamefont {J.}~\bibnamefont {Garc\'\i{}a-Bellido}}, \ and\ \bibinfo {author} {\bibfnamefont {F.}~\bibnamefont {K\"uhnel}},\ }\href {\doibase 10.1016/j.dark.2020.100755} {\bibfield  {journal} {\bibinfo  {journal} {Phys. Dark Univ.}\ }\textbf {\bibinfo {volume} {31}},\ \bibinfo {pages} {100755} (\bibinfo {year} {2021})},\ \Eprint {http://arxiv.org/abs/1906.08217} {arXiv:1906.08217 [astro-ph.CO]} \BibitemShut {NoStop}%
\bibitem [{\citenamefont {Takhistov}(2018)}]{Takhistov:2017bpt}%
  \BibitemOpen
  \bibfield  {author} {\bibinfo {author} {\bibfnamefont {V.}~\bibnamefont {Takhistov}},\ }\href {\doibase 10.1016/j.physletb.2018.05.026} {\bibfield  {journal} {\bibinfo  {journal} {Phys. Lett. B}\ }\textbf {\bibinfo {volume} {782}},\ \bibinfo {pages} {77} (\bibinfo {year} {2018})},\ \Eprint {http://arxiv.org/abs/1707.05849} {arXiv:1707.05849 [astro-ph.CO]} \BibitemShut {NoStop}%
\bibitem [{\citenamefont {Takhistov}\ \emph {et~al.}(2021)\citenamefont {Takhistov}, \citenamefont {Fuller},\ and\ \citenamefont {Kusenko}}]{Takhistov:2020vxs}%
  \BibitemOpen
  \bibfield  {author} {\bibinfo {author} {\bibfnamefont {V.}~\bibnamefont {Takhistov}}, \bibinfo {author} {\bibfnamefont {G.~M.}\ \bibnamefont {Fuller}}, \ and\ \bibinfo {author} {\bibfnamefont {A.}~\bibnamefont {Kusenko}},\ }\href {\doibase 10.1103/PhysRevLett.126.071101} {\bibfield  {journal} {\bibinfo  {journal} {Phys. Rev. Lett.}\ }\textbf {\bibinfo {volume} {126}},\ \bibinfo {pages} {071101} (\bibinfo {year} {2021})},\ \Eprint {http://arxiv.org/abs/2008.12780} {arXiv:2008.12780 [astro-ph.HE]} \BibitemShut {NoStop}%
\bibitem [{\citenamefont {Bhattacharya}\ \emph {et~al.}(2024)\citenamefont {Bhattacharya}, \citenamefont {Miller},\ and\ \citenamefont {Ray}}]{Bhattacharya:2024pmp}%
  \BibitemOpen
  \bibfield  {author} {\bibinfo {author} {\bibfnamefont {S.}~\bibnamefont {Bhattacharya}}, \bibinfo {author} {\bibfnamefont {A.~L.}\ \bibnamefont {Miller}}, \ and\ \bibinfo {author} {\bibfnamefont {A.}~\bibnamefont {Ray}},\ }\href@noop {} {\  (\bibinfo {year} {2024})},\ \Eprint {http://arxiv.org/abs/2403.13886} {arXiv:2403.13886 [hep-ph]} \BibitemShut {NoStop}%
\end{thebibliography}%

\end{document}